\documentclass[11pt]{article} 

\usepackage{packages}
\usepackage{dsfont}
\usepackage{mathrsfs}

\begin{document} 
	
\title{\bf The $r$-matrix of the Alday-Arutyunov-Frolov model}

\author{ {\bf A. Melikyan}$\,^{1}$, {\bf G. Weber}$\,^{2}$ \thanks{\tt amelik@gmail.com, gbrl.wbr@gmail.com}\\

$^1$ \sl{Instituto de F\'{\i}sica}\\
\sl{Universidade de Bras\'{\i}lia, 70910-900, Bras\'{\i}lia, DF, Brasil}\\
\sl{and}\\
\sl{{International Center of Condensed Matter Physics} }\\
\sl{C.P. 04667, Brasilia, DF, Brazil} \\

$^2$ \sl{Instituto de F\'{\i}sica}\\
\sl{Universidade de S\~{a}o Paulo, 05315-970, S\~{a}o Paulo, SP, Brasil\\}}
 
\date{\today} 

\maketitle

\begin{abstract}
	We investigate the classical integrability of the Alday-Arutyunov-Frolov model, and show that the Lax connection can be reduced to a simpler $2 \times 2$ representation. Based on this result, we calculate the algebra between the $L$-operators and find that it has a highly non-ultralocal form. We then employ and make a suitable generalization of the regularization technique proposed by Maillet for a simpler class of non-ultralocal models, and find the corresponding $r$- and  $s$-matrices. We also make a connection between the operator-regularization method proposed earlier for the quantum case, and the Maillet's symmetric limit regularization prescription used for non-ultralocal algebras in the classical theory. 
\end{abstract}

\newpage 
\tableofcontents

\section{Introduction}\label{intro}
There has been a renewed interest in the last decade to the theory of integrable models and its various techniques, due to their numerous applications in the understanding of the $AdS/CFT$ correspondence (for a comprehensive review see \cite{Beisert:2010jr,Arutyunov:2009ga}). On the other hand, this study has stimulated a deeper investigation of the subtleties associated to the quantization of integrable systems. The standard methods which work for the simpler classical integrable models \cite{Faddeev:1987ph,Faddeev:1982rn,Novikov:1984id,Korepin:1997bk
} often fail when considering a more complex theory such as the string theory on $AdS_5 \times S^5$ background, or even its smaller subsectors . 

In this paper we address one such subsector, the $\mathfrak{su}(1|1)$ subsector, and point out some of the new interesting features and problems that arise in the resulting Alday-Arutyunov-Frolov ($AAF$) model \cite{Alday:2005jm,Arutyunov:2005hd, Klose:2006dd,Callan:2004dt,Staudacher:2004tk,Alday:2005gi,Stefanski:2007dp}. The first attempt to probe its quantum integrability has been done in \cite{Melikyan:2011uf}, where it has been shown that the quantum integrability of the $AAF$ model can indeed be verified directly by checking the $S$-matrix factorization property, which is a necessary condition for the quantum integrability of the system. However, the technically complex perturbative calculation was possible to carry out only up to the 1-loop order, and it does not seem to be possible to generalize the perturbative test of this property to all-loop order. Besides, as we have emphasized in \cite{Melikyan:2011uf}, the perturbative calculations for the $AAF$ model contain several subtleties and fine points. Moreover, the most interesting properties of the integrable systems are generally non-perturbative in nature, and, therefore, developing a more strict approach is very desirable.

Thus, our goal in this paper is to develop the more reliable inverse scattering method for the $AAF$ model, which does not utilize any perturbative calculations. The standard path to quantizing an integrable system is to start from the classical integrability and the corresponding Lax connection, and find the algebra of the corresponding monodromy matrices, which determines the classical $r$-matrix of the model. The quantization is then usually achieved by putting the system onto a lattice, with the use of the classical algebra. While this works for a number of classical models, it does not work in such a straightforward manner for many interesting models, due to the following several reasons. First, while the classical algebra may be simple, finding the lattice version of the Lax pair is a very non-trivial problem. Even for the simplest classical models, for example, the sine-Gordon model, the general techniques of constructing the lattice counterparts do not give a simple local Hamiltonian, even though they may exist in principle \cite{Izergin:1982ry,Izergin:1981mc,Tarasov:1983cj,Tarasov:1984jj}. In some cases, for example, the non-linear Schr\"{o}dinger ($NLS$) model, one may write the corresponding quantum equations directly in the continuous case without any evident problems. This, however, fails for other models, as we have shown in \cite{Melikyan:2008ab,Melikyan:2010fr} for the case of the Landau-Lifshitz ($LL$) model \cite{Sklyanin:1988}, for which the direct generalization of the classical equations to the quantum ones leads to meaningless singular expressions, due to the ill-defined operator product in the same point. As was shown in \cite{Melikyan:2008ab,Melikyan:2010fr}, a more mathematically correct procedure to obtain the corresponding non-singular quantum expressions is to introduce a special operator regularization, and necessarily construct the self-adjoint extensions. Together with the self-adjointness of the quantum Hamiltonian this was shown to reproduce both the correct spectrum and the factorization property of the $S$-matrix. The root of the problem in the $LL$ model was the very singular $\delta '' (x)$ type of interaction in the quantum-mechanical description. We emphasize that even in the $NLS$ model one in principle should construct the corresponding self-adjoint extensions, although, in this case, the result is non-essential for the integrability of the system.

The $AAF$ model, which contains only fermionic degrees of freedom, is another interesting example of such singular theories. Here one also has to deal with the $\delta '' (x)$ type of potential, and, therefore, all the methods and operator regularization techniques developed for the $LL$ model could also be applied in this case. However, unlike the $LL$ model, there is a further complication in the $AAF$ model, apparent already on the classical level. As we will show below, the algebra of the $L$-operators has a non-ultralocal form, which essentially prevents one from using the standard methods of the integrable systems and quantize the model via the Bethe Ansatz techniques. Although the non-ultralocality appears in many interesting models, e.g. principal chiral mode, 2d gravity etc., the non-ultralocality in the algebra of the $L$-operators is usually exhibited in terms proportional to $\partial_{x} \delta(x-y)$. However, the corresponding algebra of the $L$-operators for the $AAF$-model has a surprisingly more complicated structure, containing higher order non-ultralocal terms of the type $\partial_{x}^{2} \delta(x-y)$. 

Thus, the $AAF$ model is an interesting model in this context, which exhibits both difficulties simultaneously - the singularity of the interaction, and the non-ultralocality of the algebra of the $L$ operators. The singular nature can be dealt with in the same manner as it was done for the $LL$ model \cite{Melikyan:2008ab,Melikyan:2010fr}. However, handling the non-ultralocal nature of the algebra is not an easy task. There do not exist any satisfactory nor standard methods to deal with such algebras. The main prescriptions in this direction are due to Maillet \cite{Maillet:1985ek,Maillet:1985fn,Maillet:1985ec,deVega:1983gy
} and, alternatively, to Faddeev and Reshetikhin \cite{Faddeev:1985qu}. The latter method, being more elegant and physically clear, is, nevertheless, hard to use in practice for more involved models, and, moreover, it still requires putting the system on the lattice.\footnote{For a recent attempt to apply this method to the strings on $AdS_5 \times S^5$  see \cite{Delduc:2012qb,Delduc:2012vq}} The method due to Maillet, however, does not use any lattice regularization, and although it is not obvious how to quantize such systems, some essential progress in understanding the integrability of such models, e.g., the complex sine-Gordon model and  non-linear sigma models, has been made in the classical theory (for more recent applications see \cite{Benichou:2012hc,Benichou:2011ch,Benichou:2010ts,Mikhailov:2007eg,Dorey:2006mx} and the references therein). 

One of the main issues regarding the method proposed by Maillet is the \emph{ad hoc} construction of the symmetric limit procedure, in order to obtain well-defined algebras between the monodromy matrices.\footnote{Remarkably, in some cases, such as the 2d gravity coupled to a dilaton field, the algebra is well-defined (in the infinite space limit), despite the non-ultralocality in the algebra of the $L$-operators, due to the presence of the dilaton field and its assymptotic behavior \cite{Korotkin:1997fi}.\label{fn:2}} In this paper we will argue that such symmetric limit procedure is the result of the regularized operator product in the quantum theory, and which naturally appears when one takes the classical limit $\hbar \rightarrow 0$. 

Another interesting result we have obtained is that the $AAF$ model admits a $2 \times 2$ Lax pair representation. Let us remind, that in \cite{Alday:2005jm} it was shown that one can obtain a $4 \times 4$ representation, starting from the Lax pair for the full superstring on $AdS_5 \times S^5$ and carefully eliminating the degrees of freedom, in order to obtain the $\mathfrak{su}(1|1)$ subsector. Moreover,  it was proved there that the the zero-curvature condition is satisfied on the equations of motion. Here we demonstrate also the inverse, namely, we show that one can derive the equations of motions from the zero-curvature condition. In the process, by analysing all the constraints and the resulting equations, we show that the Lax connections can be recast in the $2 \times 2$ matrix form, which essentially simplifies the computation of the algebra of the $L$-operators. This is quite remarkable, since the much simpler fermionic Thirring model admits only a $3 \times 3$ matrix representation for the Lax operators \cite{Kulish:1981bi,Sklyanin:1980ij}, which makes it complicated to use the Bethe Ansatz to find the spectrum. Surprisingly, the $AAF$ model appears to be simpler in this sense. 

We also give the complete account of the non-ultralocal algebra between the $L$-operators, which is hard to compute explicitly due to a very complex Dirac bracket structure between the fermionic fields. We note here, that as has been shown in \cite{Melikyan:2011uf}, in the process of checking the $S$-matrix factorization property via perturbative 1-loop calculations, we have discovered some missed numerical factors in the $AAF$ Lagrangian in the earlier works \cite{Alday:2005jm, Klose:2006dd}, which essentially changed some of the results. Here, we also correct the missed factors in the Diract bracket structure of \cite{Alday:2005jm}. We show that in order to describe the integrable structure of the $AAF$ model, one needs to introduce three independent matrices $r,s_{1}$ and $s_{2}$. The latter is due to the additional higher order non-ultralocal term in the algebra. Moreover, we derive the algebra for the transition matrices and show that it has exactly the same form as the Maillet algebra with the non-ultralocality of the simpler type, containing only the terms proportional to $\partial_{x}\delta(x-y)$. Thus, we show that the effect of the higher order non-ultralocality can be absorbed into only two independent matrices, the modified $r$ and $s$ pair. 

Our paper is organized as follows: In Section \ref{4_4_aaf_class_int}, we set up our notations and analyse the $4 \times 4$ matrix representation in details. We show that the $AAF$ equations of motion follow from the off-diagonal part of the zero-curvature condition, while the diagonal part gives some highly non-trivial constraints, satisfied on the equations of motion. In Section \ref{2_2_aaf_class_int}, we make the crucial observation that only half of the set of all equations, that follow from the zero-curvature condition, are indeed independent, and show that due to this doubling of the equations, one can reduce the Lax connections to a $2 \times 2$ matrix form. In Section \ref{sec_dirac_bracktes}, we briefly explain the Faddeev-Jackiw procedure to find the Dirac brackets, and give the corrected canonical structure between the fermionic fields. In Section \ref{sec_lax_algebra}, we explicitly calculate the algebra between the $L$ operators and show its non-ultralocal nature. As an interesting consequence, we find that the field-independent truncation of the algebra corresponds to the fermionic version of the Wadati model, which can serve as a more simple characteristic example to analyse non-ultralocal algebras. In Section \ref{sec_maillet_algebra}, we recapitulate the Maillet symmetric limit procedure to deal with such algebras, and give its generalization adapted to the more general case of an algebra containing also terms proportional to $\partial_{x}^2\delta (x-y)$. We also briefly explain the operator product regularization method proposed in \cite{Melikyan:2008ab,Melikyan:2010fr}, and argue that Maillet's symmetric limit procedure appears naturally in the classical limit of the regularized quantum case. In conclusion we discuss some open problems and the future work. Finally, we collect some important technical details in the appendices.

\section{Alday-Arutyunov-Frolov model: $4 \times 4$ Lax connection}\label{4_4_aaf_class_int}
In this section we setup our notations and give the complete analysis of the classical integrability, using the $4 \times 4$ Lax representation originally found in \cite{Alday:2005jm}. Let us note, that it was claimed there that the corresponding zero-curvature condition is satisfied upon the substitution of the equations of motion. Here we also prove the inverse:  the equations of motion of the $AAF$ model follow from the zero-curvature condition in the $4 \times 4$ representation. This is a very non-trivial result, which we present in detail below. In the process we will show that the resulting equations and constraints are such that one can reduce the Lax representation to a $2 \times 2$ form.

The $AAF$ model is obtained (for a complete analysis, see the original paper \cite{Alday:2005jm}) by starting from the full superstring theory on $AdS_5 \times S^5$ and consistently reducing it to the $\mathfrak{su}(1|1)$ subsector. The remarkable characteristic feature of this truncation is the elimination of all the bosonic degrees of freedom through the constraints. Our starting point is the Lagrangian of the $AAF$ model in the form (see appendix \ref{app_notations} for our notations): 
\begin{align}
	\label{aaf_lagrangian} \mathscr{L} &= -J - \frac{iJ}{2} \left(\bar{\psi} \rho^0 
	\partial_0 \psi - 
	\partial_0 \bar{\psi} \rho^0 \psi \right) + i \kappa \left(\bar{\psi}\rho^1 
	\partial_1 \psi - 
	\partial_1 \bar{\psi} \rho^1 \psi \right) + J\bar{\psi}\psi  \nonumber \\
	&+ \frac{\kappa g_{2}}{2} \epsilon^{\alpha \beta} \left( \bar{\psi}
	\partial_{\alpha} \psi \; \bar{\psi} \rho^5 
	\partial_{\beta} \psi -
	\partial_{\alpha}\bar{\psi} \psi \; 
	\partial_{\beta} \bar{\psi} \rho^5 \psi \right) - \frac{\kappa g_{3}}{8} \epsilon^{\alpha \beta} \left(\bar{\psi}\psi\right)^2 
	\partial_{\alpha}\bar{\psi}\rho^5 
	\partial_{\beta}\psi. 
\end{align}
Here, as explained in \cite{Melikyan:2011uf}, we have introduced two  coupling constants $g_{2}$ and $g_{3}$. The main result of the analysis in \cite{Melikyan:2011uf} was the necessary relation between the coupling constants $g_{2}$ and $g_{3}$ in order the guarantee the quantum integrability of the model. Namely, in order for the $S$-matrix factorization property to hold, up to the 1-loop order, the following relation must be satisfied:
\begin{equation}
	(g_{2})^{2}=g_{3} \label{g2g3}.
\end{equation}
We will show below that the same condition must also hold for the classically integrable theory. For now, we will consider a more general theory defined by \eqref{aaf_lagrangian} where the constants $g_{2}$ and $g_{3}$  are independent, and below we will show, that the constraint \eqref{g2g3} should be imposed already in the classical theory, from the condition of classical integrability.

To analyse the classical integrability, it is convenient to write the equations of motion, following from the Lagrangian \eqref{aaf_lagrangian}, for each component $\psi_{1}$ and $\psi_{2}$ separately:
\begin{align}
&iJ \partial_{0} \psi_{1} - \sqrt{\lambda} \partial_{1} \psi_{2} + J \psi_{1} + \frac{i \sqrt{\lambda}g_{2}}{2}\left[ -\psi_{2}^{*}\left( \partial_{0}\psi_{1} \partial_{1}\psi_{1} + \partial_{0}\psi_{2}\partial_{1}\psi_{2}\right) + \epsilon^{\alpha \beta} \partial_{\alpha} \psi_{1}^{*} \partial_{\beta}(\psi_{1} \psi_{2})\right]  \notag \\
&+ \frac{i \sqrt{\lambda}g_{3}}{8}\epsilon^{\alpha \beta}\left\{ \psi_{2}^{*} \partial_{\alpha} \psi_{2}^{*} \psi_{1} \psi_{2} \partial_{\beta} \psi_{1} 
- \left[ \partial_{\alpha} \psi_{1}^{*} \psi_{2}^{*} \psi_{1} \psi_{2} + \partial_{\alpha}(\psi_{1}^{*}\psi_{2}^{*}\psi_{1}\psi_{2})\right] \partial_{\beta}\psi_{2}\right\} =0, \label{eom_psi1} \\
&iJ \partial_{0} \psi_{2} + \sqrt{\lambda} \partial_{1} \psi_{1} - J \psi_{2} - \frac {i \sqrt{\lambda}g_{2}}{2}\left[ -\psi_{1}^{*}\left( \partial_{0}\psi_{1} \partial_{1}\psi_{1} + \partial_{0}\psi_{2}\partial_{1}\psi_{2}\right) - \epsilon^{\alpha \beta} \partial_{\alpha} \psi_{2}^{*} \partial_{\beta}(\psi_{1} \psi_{2})\right]  \notag \\
&- \frac{i \sqrt{\lambda}g_{3}}{8}\epsilon^{\alpha \beta}\left\{ \psi_{1}^{*} \partial_{\alpha} \psi_{1}^{*} \psi_{1} \psi_{2} \partial_{\beta} \psi_{2} 
+ \left[\psi_{1}^{*} \partial_{\alpha}  \psi_{2}^{*} \psi_{1} \psi_{2} + \partial_{\alpha}(\psi_{1}^{*}\psi_{2}^{*}\psi_{1}\psi_{2})\right] \partial_{\beta}\psi_{1}\right\} =0. \label{eom_psi2}
\end{align}
This form, however, is still not suitable to proceed with the analysis, since the time derivative of the spinor components enters also into the cubic and higher order terms in \eqref{eom_psi1} and \eqref{eom_psi2}. To this end, one can substitute in these higher order terms the time derivatives of the spinor components upon multiple usage of the equations \eqref{eom_psi1} and \eqref{eom_psi2}, until the higher order terms depend only on space derivatives, and the equations of motion have the form: 
\begin{equation}
\partial_{0} \psi_{i} = F_{i}(\psi_{1},\psi_{2}, \partial_{1}\psi_{1},\partial_{1}\psi_{2}).
\end{equation}
We stress that the convergence of this procedure is guaranteed by the fermionic nature of the fields. After very lengthy and tedious calculations, one obtains the expression \eqref{eom_reduced_psi1} and \eqref{eom_reduced_psi2} of appendix \ref{app_eoms_reduced}. Even though the resulting equations are rather cumbersome and have a more complicated form in comparison to the original equations \eqref{eom_psi1} and \eqref{eom_psi2}, containing terms up to the seventh order in the fermions and their space derivative, they will greatly simplify obtaining a number of very non-trivial relations, which will play a central role in establishing the classical integrability.

Let us now turn to the Lax representation. The $4 \times 4$  representation was given in \cite{Alday:2005jm}, and there it was shown that the zero-curvature condition is satisfied upon the substitution of the equations of motion. In order to establish the inverse, namely, that the zero-curvature condition produces the equations of motion \eqref{eom_psi1} and \eqref{eom_psi2}, we will first generalize the construction of \cite{Alday:2005jm}, write down the equations following from the zero-curvature condition, and find the general conditions upon which the off-diagonal terms produce the equations of motion, while the diagonal terms give some very non-trivial identities. 

The $4 \times 4$ Lax connection has the form:\footnote{As explained in \cite{Alday:2005jm}, this Lax representation was obtained by starting from the one for the full superstring on $AdS_5 \times S^{5}$ and consistently reducing it to the $\mathfrak{su}(1|1)$ subsector. We also note, that under the scale transformation: $\psi_{i} \rightarrow c \psi_{i}$, the Lagrangian \eqref{aaf_lagrangian} transforms as: $ \mathscr{L}(g_{2},g_{3}) \rightarrow c^{2}  \mathscr{L}(g_{2}', g_{3}')$, where $g_{2}' = c^{2}g_{2}$ and $g_{3}' = c^{4}g_{3}$. One can use such transformation to set: $g_{2}=1$. The Lax connection \eqref{L0} and \eqref{L1} is written after making such a scaling.\label{fn:3}}
\begin{align}
	L_{0} &= \xi_{0}^{(\tau)}I_{0} + \xi_{1}^{(\tau)}J_0 + \Lambda_{\tau}, \label{L0} \\
	L_{1} &= \xi_{0}^{(\sigma)}I_{0} + \xi_{1}^{(\sigma)}J_0 + \Lambda_{\sigma}, \label{L1}
\end{align}
where $I_0 \equiv \sigma^{3} \otimes \sigma^{3}$ and $J_0 \equiv \sigma^{3} \otimes \mathbb{1}$, and the other quantities in \eqref{L0} and \eqref{L1} are defined as follows:\footnote{We write these quantities  explicitly in terms of the spinor components $\psi_{1}$ and $\psi_{2}$ in appendix \ref{app_lax_details}.} 
\begin{align}
	\xi_{0}^{(\tau)} &= \frac{1}{4}(1+\bar{\psi}\psi)(\bar{\psi}\dot{\psi}-\dot{\bar{\psi}}\psi)+\frac{i}{2}\bar{\psi}\rho^{0}\psi \label{xi_0_tau}, \\
	\xi_{1}^{(\tau)} &=\frac{l_{1}}{8} \left(\bar{\psi}\rho^{0}\dot{\psi} - \dot{\bar{\psi}}\rho^{0}\psi + 2i\bar{\psi}\psi -4i\right)  + \frac{l_{2}\sqrt{\lambda}}{8J}\left(\bar{\psi}\rho^{0}\psi' - \bar{\psi}'\rho^{0}\psi\right),\label{xi_1_tau}\\
	\xi_{0}^{(\sigma)} &= \frac{1}{4}(1+\bar{\psi}\psi)(\bar{\psi}\psi'-\bar{\psi}'\psi), \label{xi_0_sigma} \\
	\xi_{1}^{(\sigma)} &= \frac{l_{1}}{8}\left(\bar{\psi}\rho^{0}\psi' - \bar{\psi}'\rho^{0}\psi\right) + \frac{il_{2}}{4\sqrt{\lambda}}\left[ 2J -\frac{i\sqrt{\lambda}}{2}\left(\bar{\psi}\rho^{1}\psi' - \bar{\psi}'\rho^{1}\psi \right) - J\bar{\psi}\psi \right]. \label{xi_1_sigma}
\end{align}
Furthermore, the off-diagonal matrices $\Lambda_{\tau}$ and $\Lambda_{\sigma}$ have the following form:
\begin{align}
	\Lambda_{\tau} &=\left[\gamma_{\tau}J_0, l_{3} \theta + l_{4} \tilde{\theta}\right] - \partial_{\tau}\left(l_{3} \theta - l_{4} \tilde{\theta}\right), \label{lambda_tau} \\
	\Lambda_{\sigma} &= \left[\gamma_{\sigma}J_0, l_{3} \theta + l_{4} \tilde{\theta}\right] - \partial_{\sigma}\left(l_{3} \theta - l_{4} \tilde{\theta}\right), \label{lambda_sigma}
\end{align}
where the functions $l_{i} \equiv l_{i}(\mu),$ $i=0, \ldots,4$, given explicitly in appendix \ref{app_notations}, depend on the spectral parameter $\mu$, and:
\begin{align}
	\gamma_{\sigma} &=\frac{1}{8}\left( \bar{\psi}\rho^{0}\psi' - \bar{\psi}'\rho^{0}\psi\right), \label{gamma_sigma} \\
	\gamma_{\tau} &= \frac{1}{8}\left( \bar{\psi}\rho^{0}\dot{\psi} - \dot{\bar{\psi}}\rho^{0}\psi +2i\bar{\psi}\psi - 4i \right) \label{gamma_tau}.
\end{align}
Finally, the matrices $\theta$ and $\tilde{\theta}$ have the form:
\begin{equation}
	\label{thetas} \theta = \alpha_{0}\left( 
	\begin{array}{cccc}
		0 & 0 & 0 & \nu_{1}\\
		0 & 0 & \nu_{2} & 0\\
		0 & \nu_{3} & 0 & 0\\
		\nu_{4} & 0 & 0 & 0\\
	\end{array}
	\right), \quad  \tilde{\theta} = i\alpha_{0}\left( 
	\begin{array}{cccc}
		0 & 0 & 0 & \tilde{\nu_{1}}\\
		0 & 0 & \tilde{\nu_{2}} & 0\\
		0 & \tilde{\nu_{3}} & 0 & 0\\
		\tilde{\nu_{4}} & 0 & 0 & 0\\
	\end{array}
	\right) \quad \text{with}  \quad \alpha_{0}=1 + \frac{1}{4}(\bar{\psi} \psi).
\end{equation}
The fermionic degrees $\nu_{m}, \tilde{\nu}_{m}$, $m= 1,\ldots,4$ will be determined below from the zero-curvature condition:
\begin{equation}
	\partial_{0}L_{1} - \partial_{1}L_{0} - [L_{0} ,L_{1}] = 0 \label{zcc}.
\end{equation}

In order to write the complete set of equations following from \eqref{zcc} in a more compact form, it is convenient to write the $\Lambda_{\tau}$ and $\Lambda_{\sigma}$ matrices in the basis $\sigma^{i} \otimes \sigma^{j}$, where $\sigma^{i}$ are the usual Pauli matrices. It is easy to check the following relations:
\begin{align}
	\theta= \alpha_{0}\sigma^{i} \otimes \sigma^{j} \chi_{ij} \quad \text{and} \quad \tilde{\theta}= i\alpha_{0}\sigma^{i} \otimes \sigma^{j} \tilde{\chi}_{ij}, \label{thetas_2}
\end{align}
where
\begin{align}
	\chi_{11} &= \frac{1}{4}\left( \nu_{1}+\nu_{2}+\nu_{3}+\nu_{4}\right), \quad	
	\chi_{12} = \frac{i}{4}\left( \nu_{1}-\nu_{2}+\nu_{3}-\nu_{4}\right), \nonumber\\ 
	\chi_{21} &= \frac{i}{4}\left( \nu_{1}+\nu_{2}-\nu_{3}-\nu_{4}\right), \quad 	
	\chi_{22} = \frac{1}{4}\left( -\nu_{1}+\nu_{2}-\nu_{3}+\nu_{4}\right), \label{chi_ij_1} \\
	\chi_{3i}&=\chi_{i3}=0. \nonumber
\end{align}
Similarly, one obtains the relations  between $\tilde{\chi}_{ij}$ and $\tilde{\nu}_{i}$. Finally, introducing:
\begin{align}
	\mu_{ij}^{(\pm)} = l_{3}\chi_{ij} \pm il_{4}\tilde{\chi}_{ij}, \quad \tilde{\mu}_{1i}^{(+)} = -\mu_{2i}^{(+)} \quad \text{and} \quad \tilde{\mu}_{2i}^{(+)} = -\mu_{1i}^{(+)}\label{mu_pm}
\end{align}
one arrives at the desired form:
\begin{align}
	\Lambda_{\tau} &= \Lambda_{\tau}^{ij} \sigma^{i} \otimes \sigma^{j} \quad \text{with} \quad \Lambda_{\tau}^{ij} = 2i\alpha_{0}\gamma_{\tau}\tilde{\mu}_{ij}^{(+)} - \partial_{\tau}\left( \alpha_{0}\mu_{ij}^{(-)}\right), \label{lambda_tau_tensor} \\
	\Lambda_{\sigma} &= \Lambda_{\sigma}^{ij} \sigma^{i} \otimes \sigma^{j} \quad \text{with} \quad \Lambda_{\sigma}^{ij} = 2i\alpha_{0}\gamma_{\sigma}\tilde{\mu}_{ij}^{(+)} - \partial_{\sigma}\left( \alpha_{0}\mu_{ij}^{(-)}\right). \label{lambda_sigma_tensor}
\end{align}
Writing the zero-curvature condition for the diagonal and off-diagonal parts separately, we find the following equations:
\begin{align}
	\partial_{0}\xi_{0}^{(\sigma)}I_{0} + \partial_{0}\xi_{1}^{(\sigma)}J_{0} - \partial_{1}\xi_{0}^{(\tau)}I_{0} - \partial_{1}\xi_{1}^{(\tau)}J_{0} - \left[ \Lambda_{\tau},\Lambda_{\sigma} \right] &= 0, \label{diag_part} \\
	\partial_{0}\Lambda_{\sigma} - \partial_{1}\Lambda_{\tau} - \xi_{1}^{(\tau)}\left[J_{0},\Lambda_{\sigma}\right] + \xi_{1}^{(\sigma)}\left[J_{0},\Lambda_{\tau}\right] &= 0. \label{offdiag_part}
\end{align}

We first analyse the equation \eqref{offdiag_part} for the off-diagonal part of the zero-curvature condition \eqref{zcc}, and show that the equations of motion \eqref{eom_psi1}, \eqref{eom_psi2} follow from it. Using the expressions \eqref{lambda_tau_tensor} and \eqref{lambda_sigma_tensor} for the tensors $\Lambda_{\tau}^{ij}$ and $\Lambda_{\sigma}^{ij}$, one can write the expression \eqref{offdiag_part} in the component form:
\begin{align}
	\partial_{0}\Lambda_{\sigma}^{ij} - \partial_{1}\Lambda_{\tau}^{ij} - 2i\epsilon_{3ki}\Lambda_{\sigma}^{kj} \xi_{1}^{(\tau)} + 2i\epsilon_{3ki}\Lambda_{\tau}^{kj}\xi_{1}^{(\sigma)} = 0, \quad \text{for} \quad i,j = 1,2. \label{off_diag_components}
\end{align}
Substituting all possible values for the indices $(i,j)$, and using the explicit expressions \eqref{xi_0_tau} - \eqref{xi_1_sigma}, one obtains the following system of equations:
\begin{align}
	l_{3} M_{3}^{(ij)} + l_{4} M_{4}^{(ij)}  + l_{1}l_{3} M_{13}^{(ij)} + l_{1}l_{4} M_{14}^{(ij)} + l_{2}l_{3} M_{23}^{(ij)} + l_{2}l_{4} M_{24}^{(ij)} = 0. \label{M_functions}  
\end{align}
The explicit form of the functions $M_{3}^{(ij)}, M_{4}^{(ij)}, M_{13}^{(ij)}, M_{14}^{(ij)}, M_{23}^{(ij)}$ and $M_{24}^{(ij)}$ are given in appendix \ref{app_M_functions}. By using the explicit dependence of the $l_{i}$ functions on the spectral parameter $\mu$, given in the equation \eqref{l_functions}, and analysing its various values and asymptotics, one can show that the equation \eqref{M_functions} is equivalent to the following set of constraints:
\begin{align}
	M_{3}^{(ij)} + M_{13}^{(ij)} &= 0, \label{M_3_13}\\
	M_{14}^{(ij)} - M_{23}^{(ij)} &= 0, \label{M_14_23}\\
	M_{13}^{(ij)} - M_{24}^{(ij)} &= 0,\label{M_13_24}\\
	M_{4}^{(ij)} - M_{14}^{(ij)} &= 0. \label{M_4_14}
\end{align}
Despite the complicated dependence of the $M^{(ij)}$ functions on the fields, one can show that by an appropriate choice of $\nu_{i}$ in \eqref{thetas}, the above system of equations reproduces the equations of motion \eqref{eom_psi1} and \eqref{eom_psi2}. This in turn will imply that the equations \eqref{M_3_13} - \eqref{M_4_14} are not independent, leading  to the construction of the $2 \times 2$ Lax connection, which we present in the next section.

We start with the equation \eqref{M_3_13}. Using the formulas in appendix \ref{app_M_functions}, it is easy to show that, for any choice of the $(i,j)$ indices, the equation \eqref{M_3_13} is equivalent to the following constraint:
\begin{align}
	\partial_{1} \gamma_{\tau} - \partial_{0} \gamma_{\sigma} = 0, \label{M3_M13_constraint}
\end{align}
or, more explicitly:
\begin{align}
	\psi_{1}'^{*}\dot{\psi}_{1} + \psi_{2}'^{*}\dot{\psi}_{2} - \dot{\psi}_{1}^{*}\psi_{1}' - \dot{\psi}_{2}^{*}\psi_{2}' = -i\partial_{1} \left(\bar{\psi}\psi\right), \label{M3_M13_constraint_explcit}
\end{align}
where we have used the $\gamma_{\tau}$ and $\gamma_{\sigma}$ defined correspondingly in \eqref{gamma_tau} and \eqref{gamma_sigma}. The dependence of the fields $\nu_{i},\tilde{\nu}_{i}$ on the fermionic fields must be chosen so that this constraint is satisfied. 

Let us now turn to the equation  \eqref{M_13_24}, from which the fields $\nu_{i},\tilde{\nu}_{i}$ are determined. Considering all possible choices of indices $(i,j)$, using the formulas in appendix \ref{app_M_functions}, as well as inverting the relations \eqref{chi_ij_1}, one arrives at the following equations:
\begin{align}
	2i\gamma_{\tau}\partial_{1} (\alpha_{0}\nu_{k}) &- 2i\gamma_{\sigma}\partial_{0} (\alpha_{0}\nu_{k}) - \frac{4\sqrt{\lambda}}{J}\gamma_{\sigma}\alpha_{0}\gamma_{\sigma}\tilde{\nu}_{k} -\frac{2\sqrt{\lambda}}{J}\gamma_{\sigma}\partial_{1}(\alpha_{0}\tilde{\nu}_{k}) \notag \\
	&+ \frac{i}{\sqrt{\lambda}} \zeta \alpha_{0} \gamma_{\tau} \tilde{\nu}_{k} + \frac{i}{2\sqrt{\lambda}} \zeta \partial_{0}(\alpha_{0}\tilde{\nu}_{k}) = 0, \quad \text{for} \quad  k=1,2, \label{eq_nu_1_2}
\end{align}
and
\begin{align}
	2i\gamma_{\tau}\partial_{1} (\alpha_{0}\nu_{m}) &- 2i\gamma_{\sigma}\partial_{0} (\alpha_{0}\nu_{m}) + \frac{4\sqrt{\lambda}}{J}\gamma_{\sigma}\alpha_{0}\gamma_{\sigma}\tilde{\nu}_{m} -\frac{2\sqrt{\lambda}}{J}\gamma_{\sigma}\partial_{1}(\alpha_{0}\tilde{\nu}_{m}) \notag \\
	&- \frac{i}{\sqrt{\lambda}} \zeta \alpha_{0} \gamma_{\tau} \tilde{\nu}_{m} + \frac{i}{2\sqrt{\lambda}} \zeta \partial_{0}(\alpha_{0}\tilde{\nu}_{m}) = 0, \quad \text{for} \quad m=3,4, \label{eq_nu_3_4}
\end{align}
where have we denoted:
\begin{align}
	\zeta := 2J -\frac{i\sqrt{\lambda}}{2}\left(\bar{\psi}\rho^{1}\psi' - \bar{\psi}'\rho^{1}\psi \right) - J\bar{\psi}\psi. \label{zeta_notation}
\end{align}
Thus, we see that the equations following from the off-diagonal part of the zero-curvature condition are not independent, and there are in fact only two independent equations. 

Before proceeding to the analysis of the equations \eqref{eq_nu_1_2} and \eqref{eq_nu_3_4} in all orders in $\psi_{i}$, one can first readily compare the linear terms with the ones arising from the equations of motion of the $AAF$ model \eqref{eom_psi1} and \eqref{eom_psi2}. One of the possible choices is the following:\footnote{We also note, that this choice is consistent with the involution, under which $\zeta^{*} = \zeta$, $\gamma_{\tau}^{*} = -\gamma_{\tau}$, $\gamma_{\sigma}^{*} = -\gamma_{\sigma}$ and $\alpha_{0}^{*}=\alpha_{0}$.}
\begin{align}
	\begin{array}{llll}\label{nu_aaf_choice}
	\nu_{1} = \psi_{2}, & \nu_{2} = \psi_{1}^{*}, & \nu_{3} = \psi_{1}, & \nu_{4} = \psi_{2}^{*}, \\
	\tilde{\nu}_{1} = -\psi_{1}, & \tilde{\nu}_{2} = -\psi_{2}^{*}, & \tilde{\nu}_{3} = \psi_{2}, & \tilde{\nu}_{4} = \psi_{1}^{*}.
	\end{array}
\end{align}
One may then examine the connections between the equations \eqref{M_3_13} - \eqref{M_4_14}. The explicit relations are given in details in appendix \ref{app_M_functions}, and we conclude that there are only  two independent equations, following from the off-diagonal part of the zero-curvature condition, which we take to be the following:
\begin{align}
	M_{3}^{(ij)} + M_{13}^{(ij)} &= 0, \label{M_3_13a}\\
	M_{13}^{(ij)} - M_{24}^{(ij)} &= 0.\label{M_13_24a}
\end{align} 
Substituting the formulas \eqref{xi_0_tau} - \eqref{gamma_tau} into \eqref{M_13_24a}, and considering now all orders in the fields, one can derive the dynamical equations \eqref{zcc_eom_psi1} and \eqref{zcc_eom_psi2} for ${\psi_{1}}$ and ${\psi_{2}}$ following from the off-diagonal part of the zero-curvature condition. 

One immediately sees that these equations do not seem to coincide with the equations of motion for the $AAF$ model \eqref{eom_psi1} and \eqref{eom_psi2}. This is, however, due to the presence of the time derivatives of the spinor components in the higher order terms. As we discussed earlier, one can, upon multiple substitutions of the time derivatives of the spinor components into the cubic and the higher order terms, eliminate such dependences, and arrive  at the equations of motions where all the cubic and higher order terms depend only on the fields and their space derivatives. After this very lengthy and tedious elimination procedure one arrives exactly at the equations \eqref{eom_reduced_psi1} and \eqref{eom_reduced_psi2}, provided the constraint \eqref{g2g3} is satisfied.\footnote{In terms of the rescaled fields (see footnote \ref{fn:3}), this constraint becomes: $g_{2}^{2}=g_{3}=1$.} Now, using these equations, it is simple to check the remaining independent equation of the off-diagonal part  \eqref{M_3_13a}, which, as we discussed above, is equivalent to the constraint \eqref{M3_M13_constraint_explcit}. 


Let us now consider the equations arising from the diagonal part \eqref{diag_part} of the zero-curvature condition \eqref{zcc}. Writing the $\Lambda_{\tau}$ and $\Lambda_{\sigma}$ matrices in the form:
\begin{equation}
	 \Lambda_{\tau} = \left( 
	\begin{array}{cccc}
		0 & 0 & 0 & \Lambda^{1}_{\tau}\\
		0 & 0 & \Lambda^{2}_{\tau} & 0\\
		0 & \Lambda^{3}_{\tau} & 0 & 0\\
		\Lambda^{4}_{\tau} & 0 & 0 & 0\\
	\end{array}\right) \quad \text{and} \quad
\Lambda_{\sigma} = \left(	
	\begin{array}{cccc}
		0 & 0 & 0 & \Lambda^{1}_{\sigma}\\
		0 & 0 & \Lambda^{2}_{\sigma} & 0\\
		0 & \Lambda^{3}_{\sigma} & 0 & 0\\
		\Lambda^{4}_{\sigma} & 0 & 0 & 0.
	\end{array} \right), \label{Lambda_tau_sigma_matrices} 
\end{equation}
one easily obtains the following constraints:
\begin{align}
	&\partial_{0}\xi_{1}^{(\sigma)} - \partial_{1}\xi_{1}^{(\tau)} = 0, \label{B_constraint} \\
	&\partial_{0}\xi_{0}^{(\sigma)} - \partial_{1}\xi_{0}^{(\tau)} = \frac{1}{2}\left( \phi^{11} - \phi^{22} \right),\label{A_constraint} \\
	&\phi^{11} + \phi^{22} = 0, \label{zero_anomaly}
\end{align}
where we have denoted
\begin{equation}
	\left[\Lambda_{\tau},\Lambda_{\sigma} \right] = diag(\phi^{11},\phi^{22},\phi^{33},\phi^{44}).  \label{phis}
\end{equation}
It is important to stress that, as it was the case when considering the off-diagonal part of the zero-curvature condition, each of these equations appears twice. Thus, we come to the conclusion that each independent equation that follows from the $4 \times 4$ representation of the Lax connection appears exactly twice in the full set of equations and constraints. This important observation will lead us to the construction of the $2 \times 2$ Lax connection in the next section.

The verification of  the equations \eqref{B_constraint}, \eqref{A_constraint} and \eqref{zero_anomaly} again requires very lengthy calculations, and the usage of the equations \eqref{eom_reduced_psi1} and \eqref{eom_reduced_psi2}.\footnote{We emphasize that the equations \eqref{eom_reduced_psi1} and \eqref{eom_reduced_psi2} have already been obtained from the off-diagonal part of the zero-curvature condition \eqref{zcc}, and, therefore, we are allowed to use them.} Nevertheless, it is quite remarkable that these equations are indeed satisfied. In appendix \ref{app_identities} we give some useful expressions and additional technical details related to the above constraints. 

Let us also address one  subtlety which we have so far ignored. Since the $4 \times 4$ representation for the Lax connection was obtained from the original Lax connection for the full superstring on $AdS_{5} \times S^{5}$, and, as it is well known, there is no matrix representation for the $\mathfrak{psu}(2,2|4)$ superalgebra, there is a possible anomaly in the diagonal part (for a detailed discussion, see \cite{Alday:2005gi}). In other words, the diagonal part of the zero-curvature condition for the full superstring on $AdS_{5} \times S^{5}$ should be generalized to be equal to a term of the form $\Gamma_{4 \times 4}(\psi_{i}) \mathbb{1}_4$, where $\Gamma_{4 \times 4}(\psi_{i})$ is some function depending on the fields and $\mathbb{1}_4$ is the $4 \times 4$ unit matrix. It does not vanish, in general, when reduced to smaller subsectors, which is, for example, the case for the reduction to the $AdS_{3} \times S^{3}$ subsector \cite{Alday:2005gi}. For the $AAF$ model, however, the equation \eqref{zero_anomaly} is essentially the condition that the anomaly $\Gamma_{4 \times 4}(\psi_{i})$ vanishes. Indeed, it is easy to see that the anomalous term $\Gamma_{4 \times 4}(\psi_{i})$ can be written as follows:
\begin{align}
\Gamma_{4 \times 4}(\psi_{i}) = -\frac{1}{2}\left(\phi^{11} + \phi^{22}\right). \label{anomaly_phis}
\end{align}
It is interesting to note, that for the $2 \times 2$ representation of the Lax connection, which we will give in the next section, the anomaly $\Gamma_{2 \times 2}(\psi_{i})$ identically coincides with the anomaly $\Gamma_{4 \times 4}(\psi_{i})$. Therefore,  the anomalous term vanishes in both cases. 

Finally, it is tempting to try to generalize the construction of the Lax connection \eqref{L0} - \eqref{thetas} in such a way that the classical integrability holds without imposing the quantum  constraint \eqref{g2g3}, which was originally obtained in \cite{Melikyan:2011uf} from the $S$-matrix factorization property in the 1-loop order. Without giving here the explicit analysis, which is quite tedious and follows the same type of steps we outlined above, we can state that under no deformation of the parameters, or addition of higher order terms in the formulas \eqref{xi_0_tau} - \eqref{thetas} the $AAF$ model is classically integrable for arbitrary constants $g_{2}$ and $g_{3}$. Hence, the quantum constraint \eqref{g2g3} must also be imposed in the classical theory. 

Thus, we have strictly proved the classical integrability of the $AAF$ model, based on the $4 \times 4$ Lax connection, provided the constraint \eqref{g2g3}. 

\section{$2 \times 2$ Lax connection}\label{2_2_aaf_class_int}

As we showed in the previous section, the equations that follow from the zero-curvature condition \eqref{zcc} and the $4 \times 4$ Lax connection are not independent, and each independent equation appears exactly twice in the set of all equations.  This simple observation allows one to reduce the Lax connection to a $2 \times 2$ representation. Indeed, the $2 \times 2$ Lax connection can be written in the following form:
\begin{align}
	\mathcal{L}_{0} &= \xi_{0}^{(\tau)}\bar{I}_{0} + \xi_{1}^{(\tau)}\bar{J}_0 + \bar{\Lambda}_{\tau}, \label{L0_2x2} \\
	\mathcal{L}_{1} &= \xi_{0}^{(\sigma)}\bar{I}_{0} + \xi_{1}^{(\sigma)}\bar{J}_0 + \bar{\Lambda}_{\sigma}. \label{L1_2x2}
\end{align}
Here $\xi_{0}^{(\tau)}$, $\xi_{1}^{(\tau)}$, $\xi_{0}^{(\sigma)}$ and $\xi_{1}^{(\sigma)}$ are defined by the same formulas \eqref{xi_0_tau} - \eqref{xi_1_sigma}, and the $2 \times 2$ matrices $\bar{I}_{0}$ and $\bar{J}_{0}$ have now the form:
\begin{equation}
	\bar{I}_{0} = \eta_{1}\mathbb{1}_{2} \quad \text{and} \quad \bar{J}_{0} = \eta_{2}\sigma^{3}, \label{new_I0_J0}
\end{equation}
where $\eta_{1}$ and $\eta_{2}$ are some constants which we will fix below.
The off-diagonal matrices $\bar{\Lambda}_{\tau}$ and $\bar{\Lambda}_{\sigma}$ now take the following form:
\begin{align}
	\bar{\Lambda}_{\tau} &=\left[\gamma_{\tau}\bar{J}_0, l_{3} \bar{\theta} + l_{4} \tilde{\bar{\theta}}\right] - \partial_{\tau}\left(l_{3} \bar{\theta} - l_{4} \tilde{\bar{\theta}}\right), \label{lambda_tau_2x2} \\
	\bar{\Lambda}_{\sigma} &= \left[\gamma_{\sigma}\bar{J}_0, l_{3} \bar{\theta} + l_{4} \tilde{\bar{\theta}}\right] - \partial_{\sigma}\left(l_{3} \bar{\theta} - l_{4} \tilde{\bar{\theta}}\right), \label{lambda_sigma_2x2}
\end{align}
where the matrices $\bar{\theta}$ and $\tilde{\bar{\theta}}$ are defined by:
\begin{equation}
	\label{thetas_2x2} \bar{\theta} = \alpha_{0}\left( 
	\begin{array}{cc}
		0 & \bar{\nu}_{1} \\
		\bar{\nu}_{2} & 0\\
	\end{array}
	\right) \quad \text{and} \quad \tilde{\bar{\theta}} = i\alpha_{0}\left( 
	\begin{array}{cc}
		0 & \tilde{\bar{\nu}}_{1} \\
		\tilde{\bar{\nu}}_{2} & 0\\
	\end{array}
	\right).
\end{equation}

As before, we will determine the fermionic degrees $\bar{\nu}_{m}$ and $\tilde{\bar{\nu}}_{m}$, $m= 1,2$ from the zero-curvature condition:
\begin{equation}
	\partial_{0}\mathcal{L}_{1} - \partial_{1}\mathcal{L}_{0} - [\mathcal{L}_{0} ,\mathcal{L}_{1}] = 0. \label{zcc_2x2}
\end{equation}
Introducing the notations: 
\begin{align}
	{\zeta}_{1} &= \frac{1}{2}\left(\bar{\nu}_{1} + \bar{\nu}_{2} \right), \quad {\zeta}_{2} = \frac{i}{2}\left(\bar{\nu}_{1} - \bar{\nu}_{2} \right), \label{chis_2x2} \\
	\tilde{{\zeta}}_{1} &= \frac{1}{2}\left(\tilde{\bar{\nu}}_{1} + \tilde{\bar{\nu}}_{2} \right), \quad \tilde{{\zeta}}_{2} = \frac{i}{2}\left(\tilde{\bar{\nu}}_{1} - \tilde{\bar{\nu}}_{2} \right), \label{chis__tilde_2x2}
\end{align}
from which we construct the combinations:
\begin{align}
	\bar{\mu}_{i}^{(\pm)} = l_{3}{\zeta}_{i} \pm il_{4}\tilde{{\zeta}}_{i}, \quad \tilde{\bar{\mu}}_{1}^{(+)} = -\bar{\mu}_{2}^{(+)} \quad \text{and} \quad \tilde{\bar{\mu}}_{2}^{(+)} = \bar{\mu}_{1}^{(+)}, \label{mus_tilde_2x2}
\end{align}
and writing:
\begin{align}
	\bar{\Lambda}_{\tau} &= \bar{\Lambda}_{\tau}^{i} \sigma^{i} \quad \text{with} \quad \bar{\Lambda}_{\tau}^{i} = 2i\eta_{2}\alpha_{0}\gamma_{\tau}\tilde{\bar{\mu}}_{i}^{(+)} - \partial_{\tau}\left( \alpha_{0}\bar{\mu}_{i}^{(-)}\right), \label{lambda_tau_tensor_2x2} \\
		\bar{\Lambda}_{\sigma} &= \bar{\Lambda}_{\sigma}^{i} \sigma^{i} \quad \text{with} \quad \bar{\Lambda}_{\sigma}^{i} = 2i\eta_{2}\alpha_{0}\gamma_{\sigma}\tilde{\bar{\mu}}_{i}^{(+)} - \partial_{\sigma}\left( \alpha_{0}\bar{\mu}_{i}^{(-)}\right), \label{lambda_sigma_tensor_2x2} 
\end{align}
it is easy to show that the zero-curvature condition \eqref{zcc_2x2} can be written as the equations for the diagonal and off-diagonal parts, similar to \eqref{diag_part} and \eqref{offdiag_part}:
\begin{align}
	\left( \partial_{0}\xi_{0}^{(\sigma)}  - \partial_{1}\xi_{0}^{(\tau)} \right)\bar{I}_{0} + \left( \partial_{0}\xi_{1}^{(\sigma)}  - \partial_{1}\xi_{1}^{(\tau)} \right)\bar{J}_{0} - \left[ \bar{\Lambda}_{\tau},\bar{\Lambda}_{\sigma} \right] &= 0, \label{diag_part_2x2} \\
	\partial_{0}\bar{\Lambda}_{\sigma} - \partial_{1}\bar{\Lambda}_{\tau} - \xi_{1}^{(\tau)}\left[\bar{J}_{0},\bar{\Lambda}_{\sigma}\right] + \xi_{1}^{(\sigma)}\left[\bar{J}_{0},\bar{\Lambda}_{\tau}\right] &= 0. \label{offdiag_part_2x2}
\end{align}

One can then analyse these equations similarly to the $4 \times 4$ representation case. For the off-diagonal equation \eqref{offdiag_part_2x2} one can show, repeating each step of the calculation in the previous section, that the resulting equations coincide with the equations \eqref{M_3_13a} and \eqref{M_13_24a},\footnote{We stress that the equations of motion for the $AAF$ model \eqref{eom_psi1} and \eqref{eom_psi2} follow from the equations \eqref{M_3_13a} and \eqref{M_13_24a}.} provided the constraint \eqref{g2g3} on the coupling constants $g_{2}$ and $g_{3}$, as well as the relations:
\begin{align}
	\eta_{2} = 1, \quad \bar{\nu}_{1} = \psi_{2}, \quad \bar{\nu}_{2} = \psi_{2}^{*}, \quad \tilde{\bar{\nu}}_{1} = -\psi_{1} \quad \text{and} \quad \tilde{\bar{\nu}}_{2} = \psi_{1}^{*}. \label{relations_eta_tilde_nus} 
\end{align}
To simplify the analysis of the diagonal part \eqref{diag_part_2x2}, we write:
\begin{equation}
	 \Lambda_{\tau} = \left( 
	\begin{array}{cc}
		0 &  \lambda^{1}_{\tau}\\
		\lambda^{2}_{\tau} & 0 \\
	\end{array}\right) \quad \text{and} \quad
\Lambda_{\sigma} = \left(	
	\begin{array}{cc}
			0 &  \lambda^{1}_{\sigma}\\
			\lambda^{2}_{\sigma} & 0 \\
	\end{array} \right). \label{Lambda_tau_sigma_matrices_2x2} 
\end{equation}
Noting that the second term in \eqref{diag_part_2x2} is equal to zero, due to the identity \eqref{B_constraint}, and denoting:
\begin{equation}
	\left[\bar{\Lambda}_{\tau},\bar{\Lambda}_{\sigma} \right] = diag(\phi, \phi),  \label{phis_2x2}
\end{equation}
it is easy to see that the off-diagonal part \eqref{diag_part_2x2} reduces to the following equation:
\begin{equation}
	\eta_{1} 	\left( \partial_{0}\xi_{0}^{(\sigma)}  - \partial_{1}\xi_{0}^{(\tau)} \right) - \phi = \Gamma_{2 \times 2}(\psi_{i}), \label{diag_anomaly_2x2} 
\end{equation}
where $\Gamma_{2 \times 2}(\psi_{i})$ is the term that may appear due to the possible anomaly in the diagonal part, as we discussed in the end of the previous section. Using the relations \eqref{relations_eta_tilde_nus}, which were determined from the off-diagonal part, and the identity \eqref{A_constraint}, one can show, that:
\begin{align}
	\phi &= \phi^{(11)}, \label{phi_phi11}\\
	\Gamma_{2 \times 2}(\psi_{i}) &= \frac{1}{2}\left[ \left( \eta_{1}-2 \right)\phi^{(11)} - \eta_{1}\phi^{(22)} \right]. \label{anomaly_2x2_eta1}
\end{align}
Remarkably, it follows from \eqref{anomaly_phis} and \eqref{anomaly_2x2_eta1} that the anomaly $\Gamma_{2 \times 2}(\psi_{i})$ in the $2 \times 2$ case coincides identically with the anomaly $\Gamma_{4 \times 4}(\psi_{i})$ of the $4 \times 4$ case, provided  $\eta_{1}=1$. Thus, in both cases the anomalous terms vanish.

\section{Dirac brackets}\label{sec_dirac_bracktes}

Having  derived the reduced $2 \times 2$ Lax representation for the $AAF$ model, we turn our attention to its Poisson structure. This analysis has already been carried out in the original paper \cite{Alday:2005jm}, however, due to some missed coefficients in their original Lagrangian, we carefully redo this derivation. Our first step is to obtain a Hamiltonian formulation of the model. It is convenient to rescale the fields in the Lagrangian \eqref{aaf_lagrangian} as follows: $\psi \to J^{-\frac{1}{2}} \psi$. Then, the rescaled Lagrangian becomes:\footnote{It is interesting to keep the coupling constants $g_2$ and $g_3$ independent throughout the analysis carried out in this section and, therefore, uncover the explicit dependence of the Dirac brackets on them. However, when considering the algebra of Lax operators in section \ref{sec_lax_algebra}, we shall use the rescaled fields in which the constraint is $g_2^2 = g_3 = 1$.\label{fn:dbsec}}
\begin{align}
	\label{aaf_lagrangian_rescaled} \mathscr{L} &= -J - \frac{i}{2} \left(\bar{\psi} \rho^0 
	\partial_0 \psi - 
	\partial_0 \bar{\psi} \rho^0 \psi \right) + i \frac{\sqrt{\lambda}}{2J} \left(\bar{\psi}\rho^1 
	\partial_1 \psi - 
	\partial_1 \bar{\psi} \rho^1 \psi \right) + \bar{\psi}\psi  \nonumber \\
	&+ \frac{\sqrt{\lambda} \; g_2}{4 J^2} \; \epsilon^{\alpha \beta} \left( \bar{\psi}
	\partial_{\alpha} \psi \; \bar{\psi} \rho^5 
	\partial_{\beta} \psi -
	\partial_{\alpha}\bar{\psi} \psi \; 
	\partial_{\beta} \bar{\psi} \rho^5 \psi \right) - \frac{\sqrt{\lambda} \; g_3}{16 J^3} \epsilon^{\alpha \beta} \left(\bar{\psi}\psi\right)^2 
	\partial_{\alpha}\bar{\psi}\rho^5 
	\partial_{\beta}\psi. 
\end{align}
The $AAF$ Hamiltonian can be obtained by the standard Legendre transform: 
\begin{equation}\label{legendre_transform}
	\mathscr{H} = - \Pi_{\psi} \dot{\psi} - \Pi_{\bar{\psi}} \dot{\bar{\psi}} - \mathscr{L}, 
\end{equation}
where the canonical conjugate momenta are defined by:
\begin{align}
	\Pi_{\psi} &= \frac{\partial \mathscr{L}}{\partial \dot{\psi}} = \frac{i}{2} \bar{\psi} \rho^0 + \frac{\sqrt{\lambda} \; g_2}{4 J^2} \left( - \bar{\psi} \; \bar{\psi} \rho^5 \partial_1 \psi + \bar{\psi} \partial_1 \psi \; \bar{\psi} \rho^5 \right) - \frac{\sqrt{\lambda} \; g_3}{16 J^3} \left( \bar{\psi} \psi\right)^2 \partial_1 \bar{\psi} \rho^5, \label{canonical_momenta_psi}\\
	\Pi_{\bar{\psi}} &= \frac{\partial \mathscr{L}}{\partial \dot{\bar{\psi}}} = \frac{i}{2} \rho^0 {\psi} + \frac{\sqrt{\lambda} \; g_2}{4 J^2} \left( -\psi \; \partial_1 \bar{\psi}  \rho^5 \psi + \partial_1 \bar{\psi} \psi \; \rho^5 \psi \right) - \frac{\sqrt{\lambda}\; g_3}{16 J^3} \left( \bar{\psi} \psi \right)^2 \rho^5 \partial_1 {\psi}, \label{canonical_momenta_psibar}
\end{align}	
so that the Hamiltonian becomes:
\begin{equation}\label{aaf_hamiltonian}
	\mathscr{H} = J - \frac{i \sqrt{\lambda}}{2J} \left( \bar{\psi} \rho^1 \partial_1 \psi - \partial_1 \bar{\psi} \rho^1 \psi \right) - \bar{\psi} \psi.
\end{equation}

Since the Lagrangian \eqref{aaf_lagrangian_rescaled} is linear in the time derivatives, the canonical momenta are independent of the time derivatives of the fields. Hence, the attempt to eliminate the time derivatives of the fields in favour of the canonical momenta fails, and one must analyse the constrains in the theory and construct the Dirac brackets \cite{Dirac:1964lq}. This is rather a cumbersome procedure for the $AAF$ model, which can be avoided by utilizing the equivalent prescription by Faddeev and Jackiw \cite{Faddeev:1988qp, Jackiw:1993ua}, which we briefly review below.

\subsection{Overview of the Faddeev-Jackiw formalism}

Faddeev and Jackiw's method is based on the observation that a conventional second order in time derivatives Lagrangian can always be converted to a first order in time derivatives Lagrangian by the exact same Legendre transform used to go from the Lagrangian to the Hamiltonian formulation. Let us briefly describe this approach by considering a general first order in time derivatives Lagrangian:\footnote{The Lagrangian considered in this example describes a typical mechanical system. The generalization to a field theoretical setting with the inclusion of anticommuting variables is straightforward. It will be considered in detail for the $AAF$ model in the next section.}
\begin{equation}\label{fj_lagrangian}
	L = a_i(\xi) \dot{\xi}_i - V(\xi),
\end{equation}
where $\xi_i$ denote the $2n$ phase space coordinates: 
\begin{equation*}
	\xi_i = p_i \;, \quad i = 1, \ldots, n \quad \mathrm{and} \quad \xi_i = q_i \;, \quad i = n+1, \ldots, 2n,
\end{equation*}
with the sum over repeated indices, as usual, implied, and where $a_i(\xi)$ is an arbitrary function of the $\xi_i$, without explicit time dependence. Noting the absence of first order time derivatives in the combination:
\begin{equation*}
	\frac{\partial L}{\partial \dot{\xi}_i} \dot{\xi}_i - L,
\end{equation*}
when defining a Hamiltonian by the standard Legendre transform, it is possible to identify the potential $V$ with the Hamiltonian:
\begin{equation}
	H = \frac{\partial L}{\partial \dot{\xi}_i} \dot{\xi}_i - L \equiv V.
\end{equation}
Thus the first term on the right hand side of \eqref{fj_lagrangian} defines the canonical $1$-form: $a(\xi) \equiv a_i(\xi) d\xi_i$.

The Euler-Lagrange equations obtained from \eqref{fj_lagrangian} have the form:
\begin{equation}\label{fj_eom}
	\omega_{ij}\; \dot{\xi}_j = \frac{\partial H}{\partial \xi_i}, \quad \mathrm{with} \quad \omega_{ij} = \frac{\partial a_j(\xi)}{\partial \xi_i} - \frac{\partial a_i(\xi)}{\partial \xi_j}.
\end{equation}
If the $2$-form: $\omega \equiv d a = \frac{1}{2} \omega_{ij} \; d \xi_i \; d\xi_j$ is nonsingular, then the matrix $\omega_{ij}$ is invertible, and \eqref{fj_eom} can be recast in the form:
\begin{equation}\label{fj_eom2}
	\dot{\xi}_i = \omega_{ij}^{\scriptscriptstyle{-1}} \frac{\partial V}{\partial \xi_j}.
\end{equation}  
Since $V$ is the Hamiltonian for Lagrangian \eqref{fj_lagrangian}, the equations \eqref{fj_eom2} are also Hamiltonian:
\begin{equation}\label{fj_hamiltonian_eoms}
	\dot{\xi}_i = \left\{ \xi_i , V\right\} =  \left\{ \xi_i, \xi_j \right\}\frac{\partial V}{\partial \xi_j},
\end{equation}
provided one defines the bracket such that:
\begin{equation}\label{fj_bracket}
	\left\{ \xi_i, \xi_j \right\} = \omega_{ij}^{\scriptscriptstyle{-1}}.
\end{equation}

It is important to emphasize that it was not necessary to consider any constraints in the analysis so far. They only appear in the case where the matrix $\omega_{ij}$ is singular, and a more involved analysis is required. This discussion is, however, out of the scope of the present work, since, in the case of the $AAF$ model, the matrix $\omega_{ij}$ is invertible. We refer the interested reader to \cite{Faddeev:1988qp, Jackiw:1993ua}. Finally, we stress that the bracket \eqref{fj_bracket} coincides with the one obtained through the Dirac procedure \cite{Garcia:1996ac}.

\subsection{Faddeev-Jackiw formalism for the $AAF$ model}
In this section we apply the prescription due to Faddeev and Jackiw to the $AAF$ model. We start by noting that the Lagrangian \eqref{aaf_lagrangian_rescaled} admits the following decomposition:
\begin{equation}\label{aaf_lagrangian_decomposition}
	\mathscr{L} = \mathscr{L}_{kin} - \mathscr{H},
\end{equation}
where $\mathscr{H}$ is the AAF Hamiltonian written in equation \eqref{aaf_hamiltonian} and $\mathscr{L}_{kin}$ stands for the kinetic part of the Lagrangian: 
\begin{align}\label{aaf_lagrangian_kinetic}
	\mathscr{L}_{kin} &= - \frac{i}{2} \left(\bar{\psi} \rho^0 \partial_0 \psi - \partial_0 \bar{\psi} \rho^0 \psi \right) 
	+ \frac{\sqrt{\lambda} \; g_2}{4 J^2} \; \epsilon^{\alpha \beta} \left( \bar{\psi} \partial_{\alpha} \psi \; \bar{\psi} \rho^5 
	\partial_{\beta} \psi -	\partial_{\alpha}\bar{\psi} \psi \; 	\partial_{\beta} \bar{\psi} \rho^5 \psi \right) \nonumber \\
	&- \frac{\sqrt{\lambda} \; g_3}{16 J^3} \epsilon^{\alpha \beta} \left(\bar{\psi}\psi\right)^2 \partial_{\alpha}\bar{\psi}\rho^5 
	\partial_{\beta}\psi.
\end{align}
In order to obtain an analogue of \eqref{fj_lagrangian}, we must extract the canonical $1$-form from $\mathscr{L}_{kin}$, namely, we must write:
\begin{equation}
	\mathscr{L}_{kin} = a_i \left( \chi \right) \dot{\chi}_i,
\end{equation} 
where we introduced, following the notations in \cite{Alday:2005jm}, the auxiliary notation for the fermonic fields:
\begin{equation}\label{psi_chi_fields}
	\chi_1 \equiv \psi_1\;, \quad \chi_2 \equiv \psi_2 \;, \quad \chi_3 \equiv \psi_1^* \;, \quad \chi_4 \equiv \psi_2^*\;.
\end{equation} 
In this case, the functions $a_i(\chi)$ take the form: 
\begin{align}
	a_1 &= -\frac{i}{2} \chi_3 - \frac{i \sqrt{\lambda} \; g_2}{2 J^2} \chi_3 \chi_4 \chi_1' + \frac{i \sqrt{\lambda}\; g_3}{8 J^3} \chi_1 \chi_2 \chi_3 \chi_4 \chi_4', \\
	a_2 &= -\frac{i}{2} \chi_4 - \frac{i \sqrt{\lambda} \; g_2}{2 J^2} \chi_3 \chi_4 \chi_2' + \frac{i \sqrt{\lambda}\; g_3}{8 J^3} \chi_1 \chi_2 \chi_3 \chi_4 \chi_3', \\
	a_3 &= -\frac{i}{2} \chi_1 + \frac{i \sqrt{\lambda} \; g_2}{2 J^2} \chi_1 \chi_2 \chi_3' + \frac{i \sqrt{\lambda}\; g_3}{8 J^3} \chi_1 \chi_2 \chi_3 \chi_4 \chi_2', \\
	a_4 &= -\frac{i}{2} \chi_2 + \frac{i \sqrt{\lambda} \; g_2}{2 J^2} \chi_1 \chi_2 \chi_4' + \frac{i \sqrt{\lambda}\; g_3}{8 J^3} \chi_1 \chi_2 \chi_3 \chi_4 \chi_1'.
\end{align}
The next step is to derive the Euler-Lagrange equations following from \eqref{aaf_lagrangian_decomposition}. Thus, we consider: 
\begin{equation}\label{aaf_el_variation}
	\delta \left( \mathscr{L} \right) = \delta \left[ a_i \left( \chi \right) \dot{\chi}_i \right] - \delta \left( \mathscr{H} \right) = 0,
\end{equation}
with the implied sum over $i$ going from $1$ to $4$.

It is easier to evaluate each variation separately:
\begin{align}\label{a_variation}
	\delta \left[ a_i \left( \chi \right) \dot{\chi}_i \right] 
	&= \int dy \left\{ \delta \chi_j (y) \left[ \frac{\delta a_i(x)}{\delta \chi_j(y)} - \partial_{y} \frac{\delta a_i (x)}{\delta \chi_j'(y)} \right] \dot{\chi}_i(x) \right. \nonumber\\
	 &+ \left. \delta \chi_j(x) \left[ \frac{\delta a_j (x)}{\delta \chi_i (y)} \dot{\chi}_i(y) + \frac{\delta a_j (x)}{\delta \chi_j'(y)} \dot{\chi}_{i}'(y) \right] \right\} . 
\end{align}
To make it possible to write the Euler-Lagrange equations as in \eqref{fj_eom}, we must be able to write this variation as follows:
\begin{equation}\label{a_variation_compact}
	\delta \left[ a_i \left( \chi \right) \dot{\chi}_i \right] = \delta \chi_j(x) \; \Omega_{ji}(x) \; \dot{\chi}_i (x).
\end{equation}
Clearly, the expression \eqref{a_variation} does not have such a form. However, as we will show bellow, once we fix the values of the indices $i$ and $j$, it is possible to reduce \eqref{a_variation} to \eqref{a_variation_compact}. Let us then compute the right hand side of \eqref{a_variation} for the case $i=j=1$. One has:
\begin{align}
	\int dy &\left\{ \delta \chi_1 (y) \left[ \frac{\delta a_1(x)}{\delta \chi_1(y)} - \partial_{y} \frac{\delta a_1 (x)}{\delta \chi_1'(y)} \right] \dot{\chi}_1(x) + \delta \chi_1(x) \left[ \frac{\delta a_1 (x)}{\delta \chi_1(y)} \dot{\chi}_1(y) + \frac{\delta a_1 (x)}{\delta \chi_1'(y)} \dot{\chi}_{1}'(y) \right] \right\}  \nonumber \\
	&= \delta \chi_1(x) \left[ \frac{i \sqrt{\lambda} \; g_2}{2 J^2} \left( \chi_3\chi_4' - \chi_4 \chi_3' \right) + \frac{i \sqrt{\lambda} \; g_3}{4 J^3} \chi_2 \chi_3 \chi_4 \chi_4' \right] \dot{\chi}_1 (x) \nonumber \\
	&= \delta \chi_1(x) \; \Omega_{11}(x) \; \dot{\chi}_1 (x),
\end{align}
with
\begin{equation}
	\Omega_{11}(x) = \frac{i \sqrt{\lambda} \; g_2}{2 J^2} \left( \chi_3\chi_4' - \chi_4 \chi_3' \right) + \frac{i \sqrt{\lambda} \; g_3}{4 J^3} \chi_2 \chi_3 \chi_4 \chi_4'.
\end{equation}
Repeating this calculation for the other indices $(i,j)$ we can obtain all the elements of the matrix $\Omega_{ij}(x)$, which we collect in appendix \ref{app_omega_elements}. 


Computing the variation of $\mathscr{H}$ one obtains:
\begin{align}\label{h_variation}
	\delta \mathscr{H} (x) &= \int dy \left[ \delta \chi_i (y) \; \frac{\delta \mathscr{H} (x)}{\delta \chi_i (y)} + \delta\chi_i'(y)\; \frac{\delta \mathscr{H} (x)}{\delta \chi_i'(y)} \right] = \delta \chi_i(x) \: H_i(x),
\end{align}
where we have introduced the functions:
\begin{align*}
	H_1 = \frac{\sqrt{\lambda}}{J} \chi_4' - \chi_3, \quad
	H_2 = - \frac{\sqrt{\lambda}}{J} \chi_3' + \chi_4, \quad
	H_3 = - \frac{\sqrt{\lambda}}{J} \chi_2' + \chi_1 \quad \text{and} \quad
	H_4 = \frac{\sqrt{\lambda}}{J} \chi_1' - \chi_2.
\end{align*}
Now, substituting \eqref{a_variation_compact} and \eqref{h_variation} back into the equation \eqref{aaf_el_variation}, we obtain:
\begin{align}\label{aaf_fj_eom}
	\Omega_{ij}(x) \dot{\chi}_j(x) = H_i(x),
\end{align}
which is in direct correspondence to \eqref{fj_eom}.


The Dirac structure is defined in the standard manner \cite{Faddeev:1987ph}. Let $F\left[ \chi_i(x) \right]$ and $G\left[ \chi_i(x) \right]$ be two functionals of the fields and define the Dirac brackets between them in the usual way:
\begin{equation}
	\left\{ F, G \right\} = \iint dx \; dy \; \omega^{ij} (x,y) \; \frac{\delta F}{\delta \chi_i (x)} \; \frac{\delta G}{\delta \chi_i (y)}, \label{functionals}
\end{equation}
where $\omega^{ij}$ is some function of $x,y \in \mathbb{R}$. For any even functional of the fields $F\left[ \chi_i(x) \right]$, we can write:
\begin{align}\label{pb_chi_evenf}
	\left\{\chi_k(z), F \right\} 
	&= \int dw\; \left\{ \chi_k(z), \chi_l(w) \right\} \frac{\delta F}{\delta \chi_l (w)}.
\end{align}
Clearly, the Hamiltonian \eqref{aaf_hamiltonian} is an even functional of the fields, so that we can use \eqref{pb_chi_evenf} to write Hamilton's equations as:
\begin{align}\label{aaf_hamilton_eom}
	\dot{\chi}_i(x) &= \left\{ \chi_i(x), H \right\} = \int dy \; \left\{ \chi_i(x), \mathscr{H}(y) \right\} = \iint dy\; dz\; \left\{ \chi_i(x), \chi_j(z) \right\} H_j(y) \delta(y-z) \nonumber \\ 
	&= \int dy \left\{ \chi_i(x), \chi_j(y) \right\} H_j(y).
\end{align}
Finally, since the matrix $\Omega(x)$ is non-singular,\footnote{The non-singularity of $\Omega(x)$ can be directly established by explicitly checking the existence of the inverse matrix  $\Omega^{\scriptscriptstyle{-1}}(x)$. See appendices \ref{app_omega_elements} and \ref{app_poisson_structure} for details.} one can invert it and combine the equations \eqref{aaf_fj_eom} and \eqref{aaf_hamilton_eom} as follows:
\begin{align}
	\dot{\chi}_j(x) &= \Omega^{\scriptscriptstyle{-1}}_{ij} (x) H_i (x) = \int dy\; \left\{ \chi_i(x), \chi_j(y) \right\} H_i(y).
\end{align}
Then, we obtain:
\begin{align}\label{aaf_poisson_structure}	
	\left\{ \chi_i(x), \chi_j(y) \right\} = \Omega^{\scriptscriptstyle{-1}}_{ij}(x) \delta(x-y).
\end{align}
Thus, the matrix $\Omega^{\scriptscriptstyle{-1}}(x)$ defines the Dirac structure. We collect all the matrix elements of $\Omega^{\scriptscriptstyle{-1}}(x)$ in appendix \ref{app_poisson_structure}.

\section{The Algebra of Lax operators}\label{sec_lax_algebra}
With the reduced $2\times2$ Lax connection derived in section \ref{2_2_aaf_class_int} and the Dirac brackets deduced in section \ref{sec_dirac_bracktes}, we are finally in the position to obtain the algebra between the Lax operators and show its non-ultralocal structure. We start by also rescaling the fields in the expression for the spacial component of the Lax connection \eqref{L0_2x2}: $\chi_i \to J^{-\nicefrac{1}{2}}\chi_i$, so that it is consistent with the Lagrangian \eqref{aaf_lagrangian_rescaled}, which we used to derive the Dirac algebra, provided we further impose the constraint \eqref{g2g3} by setting $g_2 =1 = g_3$.\footnote{Alternatively, we could use the inverse scaling transformations (see footnote \ref{fn:3}) to restore the general coupling constants $g_{2}$ and $g_{3}$ satisfying the constraint \eqref{g2g3}.} Moreover, we decompose the spacial component of the Lax connection in a more convenient structure:
\begin{align}\label{lax_decomposition}
	\mathcal{L}_1(x;\mu) &= \xi_{0}^{(\sigma)}(x;\mu) \bar{I}_{0} + \xi_{1}^{(\sigma)}(x;\mu) \bar{J}_0 + \bar{\Lambda}_{\sigma}(x;\mu) \nonumber \\
				  &=  \xi_{0}^{(\sigma)}(x;\mu) \mathbb{1}_2 + \xi_{1}^{(\sigma)}(x;\mu) \sigma^3+ \Lambda^{(-)}_{\sigma}(x;\mu) \sigma^+ + \Lambda^{(+)}_{\sigma}(x;\mu) \sigma^-,
\end{align}
where $\sigma^i$, $i=+,-,3$ correspond to the usual Pauli matrices. Here we also introduced the functions $\xi_{j}^{(\sigma)}(x;\mu)$, $j=0,1$ and $\Lambda_{\sigma}^{(\pm)}(x;\mu)$. The former (latter) are even (odd) polynomials of the fermionic fields, containing at most one space-derivative, the expressions of which are relegated to appendix \ref{app_lax_details}. 

Using the decomposition \eqref{lax_decomposition}, we reduce the task of computing the Dirac brackets between two L-operators  to the evaluation of the following sixteen Dirac brackets between the functions $\xi_{j}^{(\sigma)}(x;\mu)$ and $\Lambda_{\sigma}^{(\pm)}(x;\mu)$: 
\begin{align}\label{aaf_LL_algebra}
	\left\{\mathcal{L}_1(x;\mu_1) \right. & \left. \stackrel{\otimes}{,} \mathcal{L}_1(y; \mu_2) \right\} = \left\{ \xi_0^{(\sigma)} (x; \mu_1), \xi^{(\sigma)}_0 (y; \mu_2) \right\} \mathbb{1}_2 \otimes \mathbb{1}_2 
														        + \left\{ \xi_0^{(\sigma)} (x; \mu_1), \xi^{(\sigma)}_1 (y; \mu_2) \right\} \mathbb{1}_2 \otimes \sigma^3 \nonumber \\
														     &+ \left\{ \xi_0^{(\sigma)} (x; \mu_1), \Lambda^{(-)}_{\sigma} (y; \mu_2) \right\} \mathbb{1}_2 \otimes \sigma^+ 
														        + \left\{ \xi_0^{(\sigma)} (x; \mu_1), \Lambda^{(+)}_{\sigma} (y; \mu_2) \right\} \mathbb{1}_2 \otimes \sigma^- \nonumber  \displaybreak[3]\\
														     &+ \left\{ \xi_1^{(\sigma)} (x; \mu_1), \xi^{(\sigma)}_0 (y; \mu_2) \right\} \sigma^3 \otimes \mathbb{1}_2 
														       +  \left\{ \xi_1^{(\sigma)} (x; \mu_1), \xi^{(\sigma)}_1 (y; \mu_2) \right\} \sigma^3 \otimes \sigma^3 \nonumber \\
														     &+  \left\{ \xi_1^{(\sigma)} (x; \mu_1), \Lambda^{(-)}_{\sigma} (y; \mu_2) \right\} \sigma^3 \otimes \sigma^+ 
														       +  \left\{ \xi_1^{(\sigma)} (x; \mu_1), \Lambda^{(+)}_{\sigma} (y; \mu_2) \right\} \sigma^3 \otimes \sigma^- \nonumber \displaybreak[3] \\
														     &+  \left\{ \Lambda^{(-)}_{\sigma} (x; \mu_1), \xi^{(\sigma)}_0 (y; \mu_2) \right\} \sigma^+ \otimes \mathbb{1}_2 
														       +  \left\{ \Lambda^{(-)}_{\sigma} (x; \mu_1), \xi^{(\sigma)}_1 (y; \mu_2) \right\} \sigma^+ \otimes \sigma^3 \nonumber \\
														    &+  \left\{ \Lambda^{(-)}_{\sigma} (x; \mu_1), \Lambda^{(-)}_{\sigma} (y; \mu_2) \right\} \sigma^+ \otimes \sigma^+ 
														    +  \left\{ \Lambda^{(-)}_{\sigma} (x; \mu_1), \Lambda^{(+)}_{\sigma} (y; \mu_2) \right\} \sigma^+ \otimes \sigma^- \nonumber  \displaybreak[3]\\
														    &+ \left\{ \Lambda^{(+)}_{\sigma} (x; \mu_1), \xi^{(\sigma)}_0 (y; \mu_2) \right\} \sigma^- \otimes \mathbb{1}_2 
														      + \left\{ \Lambda^{(+)}_{\sigma} (x; \mu_1), \xi^{(\sigma)}_1 (y; \mu_2) \right\} \sigma^- \otimes \sigma^3 \nonumber\\
														    &+  \left\{ \Lambda^{(+)}_{\sigma} (x; \mu_1), \Lambda^{(-)}_{\sigma} (y; \mu_2) \right\} \sigma^- \otimes \sigma^+
														    +  \left\{ \Lambda^{(+)}_{\sigma} (x; \mu_1), \Lambda^{(+)}_{\sigma} (y; \mu_2) \right\} \sigma^- \otimes \sigma^-.   
\end{align}
At first glance it seems that this decomposition only makes the calculations even more daunting. However, this computation is, in fact, easier, since only half of the Dirac brackets need to be evaluated. The other brackets can be obtained by taking the involution  of the corresponding brackets, as we will explain bellow.

Let $A\left(\chi_i(x)\right)$ and $B\left( \chi_i(y)\right)$ be two arbitrary functions of the fields, then the behaviour of the Dirac brackets defined by the matrix  $\Omega^{\scriptscriptstyle{-1}}$ through \eqref{aaf_poisson_structure} under involution is determined by the parity of such functions,
\begin{equation}
\left\{ A(x), B(y) \right\}^* = \begin{dcases}
	+ \left\{ A^*(x), B^*(y) \right\}, & \text{ if } A \text{ or } B \text{ even }, \\
	- \left\{ A^*(x), B^*(y) \right\}, & \text{ if } A \text{ and } B \text{ odd }.
\end{dcases}
\end{equation}
Furthermore, taking into account that the functions $\xi_{j}^{(\sigma)}(x;\mu)$ and $\Lambda_{\sigma}^{(\pm)}(x;\mu)$ have a defined parity and behave under involution as:
\begin{equation}
	{\xi_i^{(\sigma)}}^*(x,\mu) = - \xi_i^{(\sigma)}(x,\mu^*) \quad \mathrm{and} \quad {\Lambda^{(-)}_{\sigma}}^*(x,\mu) = {\Lambda^{(+)}_{\sigma}}(x,\mu^*),
\end{equation}
one can obtain some non-trivial relations amongst the following brackets:
\begin{align}
	\left\{ \xi_i^{(\sigma)}(x;\mu_1), \Lambda^{(\pm)}_{\sigma}(y,\mu_2) \right\}^* &= - \left\{ \xi_i^{(\sigma)}(x;\mu_1^*), \Lambda^{(\mp)}_{\sigma}(y,\mu_2^*) \right\}, \label{pb_involution_1} \\
	\left\{ {\Lambda^{(\pm)}_{\sigma}}(x,\mu_1),  {\Lambda^{(\pm)}_{\sigma}}(y,\mu_2) \right\}^* &= - \left\{ {\Lambda^{(\mp)}_{\sigma}}(x,\mu_1^*),  {\Lambda^{(\mp)}_{\sigma}}(y,\mu_2^*) \right\} \label{pb_involution_2}. 
\end{align}

Moreover, if we restrict the arbitrary functions of the fields $A\left(\chi_i(x)\right)$ and $B\left( \chi_i(y)\right)$ to the subset comprised of the functions $\xi_{j}^{(\sigma)}(x;\mu)$ and $\Lambda_{\sigma}^{(\pm)}(x;\mu)$, one can show that the Dirac brackets between $A\left(\chi_i(x)\right)$ and $B\left( \chi_i(y)\right)$ have the general form:\footnote{The relevant question one could still pose about the generality of the expression \eqref{pb_general_form} regards the truncation of this series at the second derivative of the delta function. As a matter of fact, there cannot be terms proportional to the third derivative of the delta function or higher, because the functions $\xi_{j}^{(\sigma)}(x;\mu)$ and $\Lambda_{\sigma}^{(\pm)}(x;\mu)$ are at most linear in the space derivatives. Hence, at most two space derivatives can act on the delta function after the Dirac brackets are computed. We note, nevertheless, that the functions $f_i(x)$ can still be proportional to some derivative of the fields. In this case the derivatives may come not only from the functions $\xi_{j}^{(\sigma)}(x;\mu)$ and $\Lambda_{\sigma}^{(\pm)}(x;\mu)$, but also from the Dirac structure itself.}
\begin{equation}\label{pb_general_form}
	\left\{ A(x), B(y) \right\} = f_1(x) \delta(x - y) + f_2(x) \partial_x \delta(x-y) + f_3(x) \partial_x^2 \delta(x-y),
\end{equation}
where $f_i(x)$, $i=1,2,3$ are some polynomials of the fields and their space derivatives. Then, by invoking the (anti)symmetry of the brackets: 
\begin{equation}
\left\{ A(x), B(y) \right\} = \begin{dcases}
	- \left\{B(y), A(y) \right\}, & \text{ if } A \text{ or } B \text{ even }, \\
	+ \left\{ B(y), A(x) \right\}, & \text{ if } A \text{ and } B \text{ odd },
\end{dcases}
\end{equation}
one can easily derive the following relation:
\begin{align}\label{pb_general_form_inverted}
	\left\{ B(x), A(y) \right\} &= \mp \left\{ \left[ f_1(x) - \partial_x f_2(x) + \partial_x^2 f_3(x) \right] \delta(x-y) + \left[- f_2(x) + 2 \partial_x f_3(x) \right]\partial_x \delta(x-y) \right. \nonumber \\
	&+ \left.  f_3(x) \partial_x^2\delta(x-y) \right\},
\end{align}
where the $f_i(x)$ are the same functions appearing in \eqref{pb_general_form}. Here the plus sign corresponds to the case where both fields are odd and the minus sign to the case where at least one of the fields is even. Hence, even though there is no simple equation that relates $\left\{ A(x), B(y) \right\}$ to $\left\{ B(x), A(y) \right\}$, we have  showed that by knowing the former one can readily obtain the expression for the latter. 

By using the properties described by the equations \eqref{pb_involution_1} and \eqref{pb_involution_2}, and by the relations \eqref{pb_general_form} and \eqref{pb_general_form_inverted}, we can drastically reduce the amount of calculations necessary to derive the algebra of the Lax operators. Namely, out of the sixteen brackets in equation \eqref{aaf_LL_algebra}, only seven are actually independent. Besides that, these properties are also extremely useful for computing each of these independent brackets. 

Before proceeding with the calculation, it is worth noting that both \eqref{pb_general_form} and \eqref{pb_general_form_inverted} already display a severe non-ultralocality, as both of them, not only contain the first derivative of the delta function, but also its second derivative. Given the structure of \eqref{aaf_LL_algebra}, one could still hope that some non-trivial cancelations would still render the algebra ultralocal, or at least diminish its non-ultralocality. However, this is not the case for the AAF model, and, as we will show bellow, they remain even after all the terms are added together.



Computing all the independent brackets in \eqref{aaf_LL_algebra} is still a very lengthy and tedious task. We sketch here only the main steps of this computation for one of the independent brackets. For the sake of clarity, let us consider the first brackets in \eqref{aaf_LL_algebra}. We first decompose it in a sum of Dirac brackets such that each entry of the Dirac brackets contains only monomials of the fields at some given order. For instance,
\begin{align}\label{pb_A}
	\left\{ \xi_0^{(\sigma)} (x, \mu_1), \xi^{(\sigma)}_0 (y, \mu_2) \right\} = \frac{1}{16} &\bigg[ \frac{1}{J^2} \left\{ A_1(x;\mu_1), A_1(y;\mu_2) \right\} + \frac{1}{J^3}\left\{ A_1(x;\mu_1), A_2(y;\mu_2) \right\} \bigg. \nonumber \\
		&+ \bigg. \frac{1}{J^3} \left\{ A_2(x;\mu_1), A_1(y;\mu_2) \right\} + \frac{1}{J^4}\left\{ A_2(x;\mu_1), A_2(y;\mu_2) \right\} \bigg],
\end{align}
where $A_1$ contains only quadratic terms in the fields and its derivatives, and $A_2$ only quartic:
\begin{align}
	A_1&=- \chi_3\chi_1' + \chi_4\chi_2' - \chi_1\chi_3' + \chi_2\chi_4', \\ 
	A_2&= \chi_2\chi_3\chi_4\chi_1' + \chi_1\chi_3\chi_4\chi_2' + \chi_1\chi_2\chi_4\chi_3' - \chi_1\chi_2\chi_3\chi_4'.
\end{align}

It is easy, though tiresome, to compute a complete set of all the Dirac brackets between all possible combinations of monomials of the fields and its derivatives. For example, the one needed to compute $\left\{ A_1(x;\mu_1), A_1(y;\mu_2) \right\}$ is:
\begin{align}
	\left\{ \chi_i(x) \chi_j'(x), \chi_k(y) \chi_l'(y) \right\} &= \left\{ \partial_x \left[ \chi_i \chi_k' \Omega^{\scriptscriptstyle{-1}}_{jl} - \chi_i \chi_l' \Omega^{\scriptscriptstyle{-1}}_{jk} \right]  \right. \nonumber \\
	&- \left. \chi_i' \chi_k' \Omega^{\scriptscriptstyle{-1}}_{jl} - \chi_i' \chi_l' \Omega^{\scriptscriptstyle{-1}}_{jk} - \chi_j' \chi_k' \Omega^{\scriptscriptstyle{-1}}_{il} - \chi_j' \chi_l' \Omega^{\scriptscriptstyle{-1}}_{ik} \right\} \delta(x-y) \nonumber \\
	&+ \left\{ \partial_x \left[ \chi_i \chi_k \Omega^{\scriptscriptstyle{-1}}_{jl} \right] + \left[ \chi_k\chi_i' + \chi_i \chi_k'\right] \Omega_{jl}^{\scriptscriptstyle{-1}} + \chi_i \chi_l' \Omega^{\scriptscriptstyle{-1}}_{jk} + \chi_k \chi_j' \Omega^{\scriptscriptstyle{-1}}_{il} \right\} \partial_x \delta(x-y) \nonumber \\
	&+ \chi_i \chi_k \Omega^{\scriptscriptstyle{-1}}_{jl} \partial_x^2 \delta(x-y).
\end{align}
After multiple substitutions and  summations one obtains the expressions of the type:
\begin{align}
	\left\{ A_1(x;\mu_1), A_1(y;\mu_2) \right\} &= \frac{1}{16} \left[ 2 \partial_x \Gamma_{11} \delta(x-y) + 4 \Gamma_{11} \partial_x \delta(x-y) \right], \label{pb_a1a1}\\
	 \left\{ A_1(x;\mu_1), A_2(y;\mu_2) \right\} &= \frac{1}{2} \left[ \Gamma_{12}^{(1)} \delta(x-y) + \Gamma_{12}^{(2)} \partial_x \delta(x-y) \right], \label{pb_a1a2}
\end{align}
where the explicit expressions for $\Gamma_{11}$, $\Gamma_{12}^{(1)}$ and $\Gamma_{12}^{(2)}$ are presented in appendix \ref{app_lax_details}. 

Deriving the expression for $ \left\{ A_2(x;\mu_1), A_1(y;\mu_2) \right\}$ becomes a straightforward task by employing the property displayed by equations \eqref{pb_general_form} and \eqref{pb_general_form_inverted}, since we can read off the concrete formulas for the $f_i(x)$ by comparing \eqref{pb_general_form} to \eqref{pb_a1a2} and then simply substitute them into \eqref{pb_general_form_inverted} to obtain: 
\begin{equation}\label{pb_a2a1}
	 \left\{ A_2(x;\mu_1), A_1(y;\mu_2) \right\} = \frac{1}{2} \left\{ \left[ -\Gamma_{12}^{(1)} + \partial_x \Gamma_{12}^{(2)}\right] \delta(x-y) + \Gamma_{12}^{(2)} \partial_x \delta(x-y) \right\}.
\end{equation}
Finally, we consider the last brackets in \eqref{pb_A}. This calculation becomes rather simple, by realizing that the relevant Dirac brackets have the form:
\begin{equation}
	\left\{ \chi_i(x) \chi_j(x) \chi_l(x) \chi_m'(x), \chi_n(y)\chi_p(y)\chi_q(y)\chi_r'(y)\right\} = f_1(x) \delta(x-y).
\end{equation}
Thus,
\begin{equation}\label{pb_a2a2_1}
	\left\{ A_2(x;\mu_1), A_2(y;\mu_2) \right\} = F(x) \delta(x-y),
\end{equation}
where both $f_1(x)$ and $F(x)$ are some polynomials in the fields and their spacial derivatives. Since $A_2$ is an even function of the fermionic fields, we can use property outlined by the equations \eqref{pb_general_form} and \eqref{pb_general_form_inverted} to write: 
\begin{equation}\label{pb_a2a2_2}
	\left\{ A_2(y;\mu_1), A_2(x;\mu_1) \right\} = - F(x) \delta(x-y).
\end{equation}
On the other hand, we can use the fact that $A_2$ does not depend on the spectral parameter, and one can simply exchange $x \leftrightarrow y$ in the equation \eqref{pb_a2a2_1} to conclude that:
\begin{equation}\label{pb_a2a2}
	\left\{ A_2(x;\mu_1), A_2(y;\mu_2) \right\} = 0.
\end{equation}

It is interesting to stress that the explicit form of the matrix elements $\Omega_{ij}^{\scriptscriptstyle{-1}}$ was not used in the derivation of \eqref{pb_a1a1}, \eqref{pb_a1a2}, \eqref{pb_a2a1} nor \eqref{pb_a2a2}, that is, $\Gamma_{11}$, $\Gamma_{12}^{(1)}$ and $\Gamma_{12}^{(2)}$ are still written in terms of $\Omega_{ij}^{\scriptscriptstyle{-1}}$. This means that the non-ultralocal form of the decompositions \eqref{pb_a1a1}, \eqref{pb_a1a2} and \eqref{pb_a2a1} are not a direct consequence of the Dirac structure of the model, but, up to this point, only of the form of the Lax connection. Thus, were $\Gamma_{11}$ and $\Gamma_{12}^{(2)}$ to vanish identically upon the substitutions of the expressions for $\Omega_{ij}^{\scriptscriptstyle{-1}}$, the equations \eqref{pb_a1a1}, \eqref{pb_a1a2} and \eqref{pb_a2a1} would be ultralocal. However, this is not the case. In fact, after careful substitution of the expressions for $\Omega_{ij}^{\scriptscriptstyle{-1}}$, given in appendix \ref{app_poisson_structure}, we conclude that:
\begin{equation}\label{pb_A_final}
	\left\{ \xi_0^{(\sigma)} (x, \mu_1), \xi^{(\sigma)}_0 (y, \mu_2) \right\} = \frac{1}{16J^2} \left\{ \partial_x \left[ 2 \Gamma_{11} + \frac{\Gamma_{12}^{(2)}}{J} \right] \delta(x-y) + 2 \left[ 2 \Gamma_{11} + \frac{\Gamma_{12}^{(2)}}{J} \right] \partial_x \delta(x-y) \right\}.
\end{equation}
where the sum $2 \Gamma_{11} + \frac{1}{J}\Gamma_{12}^{(2)}$ extends up to the sixth order in the fermion and its spacial derivatives. We relegate this lengthy expression to appendix \ref{app_lax_details}.

The remaining Dirac brackets in \eqref{aaf_LL_algebra} can be computed in exactly the same vein. In fact, the brackets which we have just computed are one of the simplest. Some of them are even plagued by more severe non-ultralocality, which extends up to the second derivative of the delta function, in full consonance with what one would  expect from \eqref{pb_general_form}. It turns out that even after summing all the brackets according to \eqref{aaf_LL_algebra}, this severe non-ultralocality survives. In particular, this prevents one from using the standard methods in the context of integrable models, as it is not possible, in general, to introduce a well-defined $r$-matrix formulation starting from a non-ultralocal algebra.\footnote{There are, however, some exceptions for which the non-ultralocal algebra for $L$-operators still leads to a well-defined algebra for the monodromy matrices. See the footnote \ref{fn:2}.} 
Nevertheless, there exists a general framework to deal with such non-ultralocal systems, which was introduced by Maillet in \cite{Maillet:1985ek,Maillet:1985fn,Maillet:1985ec,deVega:1983gy
},\footnote{See also the earlier attempts by Tsyplaev \cite{Tsyplyaev:1981cz}.} generalization of which, suitable for the  $AAF$-model, we consider in section \ref{sec_maillet_algebra}.

\subsection{Fermionic Wadati Model}

Although the $AAF$ model provides a very representative and interesting example of highly non-ultralocal models, its algebra is by far too complicated for an initial analysis. Therefore, it is more illuminating to consider first a simpler model. One such model can be obtained by considering the field independent truncation of the $AAF$ Dirac algebra, which, as we show below, corresponds to the fermionic version of the Wadati model \cite{wadati:1980ts,Tsyplyaev:1981cz}. In this case, the intricate Dirac structure $\Omega^{\scriptscriptstyle{-1}}_{ij}$ presented in appendix \ref{app_poisson_structure} simplifies considerably, and the resulting Dirac brackets are the canonical ones:
\begin{align}\label{canonical_pb}
	\begin{array}{l}
	\left\{ \chi_1(x), \chi_3(y) \right\} = i \delta(x-y), \\
	\left\{ \chi_2(x), \chi_4(y) \right\} = i \delta(x-y).
	\end{array}
\end{align}

However, despite the simplicity of the Dirac brackets, the algebra between the Lax operators \eqref{aaf_LL_algebra} is still quite complicated, as a result of the form of the Lax connection itself.\footnote{See the discussion bellow \eqref{pb_a2a2}.} Substitution of the canonical Dirac structure \eqref{canonical_pb} into the independent brackets in \eqref{aaf_LL_algebra} corresponds to their truncation to the first order in the fermion:
\begin{align} 
\left\{\xi^{(\sigma)}_0 (x;\mu_1) \right.&, \left. \Lambda^{(+)}_{\sigma}(y;\mu_2)  \right\} = \frac{1}{4J^{\frac{3}{2}}} \left[-i l_3(\mu_2) \chi_4' + l_4(\mu_2) \chi_3' \right] \partial_x \delta(x-y) \\
	&+ \frac{1}{4J^{\frac{3}{2}}}\left[ i l_3(\mu_2) \chi_4 - l_4(\mu_2) \chi_3 \right] \partial_x^2 \delta(x-y), \nonumber \\
	\left\{ \xi^{(\sigma)}_1 (x;\mu_1) \right.&, \left. \Lambda^{(+)}_{\sigma}(y;\mu_2)  \right\} = \frac{1}{8J^{\frac{3}{2}}}  \Bigg( \frac{2J}{\sqrt{\lambda}} \left[ i l_2(\mu_1) l_4(\mu_2) \chi_3 + l_2(\mu_1)l_3(\mu_2) \chi_4 \right] \\
	&- \left[l_1(\mu_1)l_4(\mu_2) + l_2(\mu_1)l_3(\mu_2) \right] \chi_3' - i \left[ l_1(\mu_1)l_3(\mu_2) + l_2(\mu_1)l_4(\mu_2) \right] \chi_4' \Bigg) \partial_x \delta(x-y) \nonumber \\
	+ \frac{1}{8J^{\frac{3}{2}}} &\Bigg( \left[ l_1(\mu_1)l_4(\mu_2) + l_2(\mu_1)l_3(\mu_2) \right] \chi_3 + i \left[ l_1(\mu_1)l_3(\mu_2) + l_2(\mu_1)l_4(\mu_2)\right]\chi_4 \Bigg) \partial_x^2 \delta(x-y), 	\nonumber \\	\left\{ \Lambda^{(-)}_{\sigma} (x;\mu_1) \right.&, \left. \Lambda^{(+)}_{\sigma}(y;\mu_2)  \right\} = -\frac{i}{J^2} \left[ l_3(\mu_1)l_3(\mu_2) + i l_4(\mu_1)l_4(\mu_2) \right] \partial_x^2 \delta(x-y).
\end{align}
The remaining brackets from \eqref{aaf_LL_algebra} are either identically zero at this order, or can be trivially derived through the use of properties \eqref{pb_involution_1}, \eqref{pb_involution_2},  \eqref{pb_general_form} and \eqref{pb_general_form_inverted}. 

Collecting everything together, we can write the algebra between the Lax operators as follows:
\begin{align}\label{wadati_LL_algebra}
	\left\{\mathcal{L}_1(x;\mu_1) \stackrel{\otimes}{,} \mathcal{L}_1(y; \mu_2) \right\} &= N_0(x,y;\mu_1,\mu_2) \delta(x-y) + N_1(x,y;\mu_1,\mu_2) \partial_x\delta(x-y) \nonumber \\ 
	&+ N_2(x,y;\mu_1,\mu_2) \partial_x^2\delta(x-y),
\end{align}
with the matrices:
\begin{align}
	N_0(x,y;\mu_1,\mu_2) &= \frac{1}{8} \left( \begin{array}{cccc}
		0 & 0 & - {N_0^{(1)}}^*(\mu_1^*,\mu_2^*) & 0 \\
		0 & 0 & 0& - {N_0^{(2)}}^*(\mu_1^*,\mu_2^*) \\
		N_0^{(1)}(\mu_1,\mu_2) & 0 & 0 & 0\\
		0 & N_0^{(2)} (\mu_1,\mu_2) & 0& 0
		\end{array} \right), \displaybreak[3]\\
	N_1(x,y;\mu_1,\mu_2) &= \frac{1}{8} \left( \begin{array}{cccc}
		0 & - {N_1^{(1)}}^*(\mu_1^*,\mu_2^*) & - {N_1^{(2)}}^*(\mu_1^*,\mu_2^*) & 0 \\
		{N_1^{(1)}}(\mu_1,\mu_2) & 0 & 0 & - {N_1^{(3)}}^*(\mu_1^*,\mu_2^*)\\
		{N_1^{(2)}}(\mu_1,\mu_2) & 0 & 0 & - {N_1^{(4)}}^*(\mu_1^*,\mu_2^*)\\
		0& {N_1^{(3)}}(\mu_1,\mu_2) & {N_1^{(4)}}(\mu_1,\mu_2) & 0
		\end{array} \right), \displaybreak[3]\\
	N_2(x,y;\mu_1,\mu_2) &= \frac{1}{8} \left( \begin{array}{cccc}
		0 & - {N_2^{(1)}}^*(\mu_1^*,\mu_2^*)  & - {N_2^{(2)}}^*(\mu_1^*,\mu_2^*) & 0 \\
		{N_2^{(1)}}(\mu_1,\mu_2) & 0 & {N_2^{(5)}}(\mu_1,\mu_2)  & - {N_2^{(3)}}^*(\mu_1^*,\mu_2^*)  \\
		{N_2^{(2)}}(\mu_1,\mu_2) & {N_2^{(5)}}(\mu_1,\mu_2) & 0 & - {N_2^{(4)}}^*(\mu_1^*,\mu_2^*)  \\
		0 & {N_2^{(3)}}(\mu_1,\mu_2) & {N_2^{(4)}}(\mu_1,\mu_2) & 0 
		\end{array} \right).
\end{align}
The functions $N_i^{(j)}(\mu_1,\mu_2)$ are defined in appendix \ref{app_N_functions}. 

Thus, restricting to the field-independent part in the algebra \eqref{wadati_LL_algebra}, we find:
\begin{align}
	\left\{\mathcal{L}_1(x; \beta_{1}) \stackrel{\otimes}{,} \mathcal{L}_1(y; \beta_{2}) \right\} = -\frac{i}{J}\cosh{(\beta_{1}+\beta_{2})}\left[ \sigma^{+} \otimes \sigma^{-} + \sigma^{-} \otimes \sigma^{+}\right] \partial_x^2\delta(x-y), \label{fermionic_wadati}
\end{align}
where we have used the parametrization \eqref{alternative_l_functions}  for the $l_{i}$  functions. The algebra \eqref{fermionic_wadati} has, curiously, the same structure as the Wadati model, and therefore, can be considered its fermionic counterpart. We refer to the original literature \cite{wadati:1980ts,Tsyplyaev:1981cz} for more details and discussions of the Wadati model. We only mention here, that despite the highly non-ultralocal algebra, the classical $r$-matrix in the infinite space limit, nevertheless, can be found. It is given, for example, in \cite{Tsyplyaev:1981cz}. Although the full algebra for the $AAF$-model is considerably more complicated than the truncated algebra \eqref{fermionic_wadati}, it is still an interesting example to consider, as it demonstrates the necessity of new methods without technical complications. 

\section{Generalized Maillet algebra} \label{sec_maillet_algebra}

In this section we will briefly explain the Maillet's formalism of the $r$- and $s$-matrices, and the symmetric limit prescription  to deal with the non-ultralocalities in the algebra. Afterwards we shall develop its generalization suitable for the $AAF$-model. The latter is needed, since, as we have seen in the previous section, the algebra for the $AAF$ model contains terms up to the second derivative of the delta function. The more complete account of such generalized Maillet algebras will be given in the separate publication \cite{melikyan:2013toappear}.

The starting point of the  formalism considered by Maillet is the non-ultralocal algebra of the form:\footnote{The splitting into $B(z,z';\lambda,\mu)$ and $C(z,z';\lambda,\mu)$ functions is introduced for the sake of convenience. One may choose these functions (as it was done for example in the original paper \cite{Maillet:1985ek}), so that  $\mathbb{P}B(z,z';\lambda,\mu)\mathbb{P} = -C(z',z;\mu,\lambda)$, where $\mathbb{P}$ is the permutation operator. This, however, is not necessary, and the choice can be a different one.}
	\begin{align}
		\{ \mathcal{L}(z;\lambda) \overset{\otimes}{,} \mathcal{L}(z';\mu) \} = A(z,z';\lambda) \delta(z-z') + B(z,z';\lambda) \partial_{z'}\delta(z-z') + C(z,z';\lambda) \partial_{z}\delta(z-z'), \label{maillet_original_alg}
	\end{align}
which can be recast in the following more convenient form:
	\begin{align}
				\{ \mathcal{L}(z;\lambda) \overset{\otimes}{,} \mathcal{L}(z';\mu) \} &= \left( \partial_{z}r(z;\lambda,\mu) + \left[r(z;\lambda,\mu),\mathcal{L}(z;\lambda) \otimes \ \mathbb{1} + \mathbb{1} \otimes \mathcal{L}(z;\mu) \right] \right. \notag \\
				&\left. +  \left[s(z;\lambda,\mu),\mathbb{1} \otimes \mathcal{L}(z;\mu) - \mathcal{L}(z;\lambda) \otimes \ \mathbb{1} \right] \right)\delta(z-z') \notag \\
				&+s(z;\lambda,\mu)\partial_{z'}\delta(z-z') + s(z';\lambda,\mu)\partial_{z}\delta(z-z').  \label{maillet_r_s_matrices}
		\end{align}
The $r$- and the new $s$-matrices are defined as follows:
\begin{align}
			&s(z;\lambda,\mu) = \frac{1}{2} \left[B(z,z;\lambda,\mu) - C(z,z;\lambda,\mu) \right], \label{maillet_s_matrix} \\
			&r(z;\lambda,\mu) = \frac{1}{2} \left[B(z,z;\lambda,\mu) + C(z,z;\lambda,\mu) \right] + r_{0}(z;\lambda,\mu), \label{maillet_r_matrix}
		\end{align}
with the function $r_{0}(z;\lambda,\mu)$ in \eqref{maillet_r_matrix} being determined from the equation:
\begin{align}
	\partial_{z}r_{0}(z;\lambda,\mu) + \left[r_{0}(z;\lambda,\mu),\mathcal{L}(z;\lambda) \otimes \ \mathbb{1} + \mathbb{1} \otimes \mathcal{L}(z,\mu) \right] = \Omega(z;\lambda,\mu), \label{maillet_r0_matrix}
\end{align}
where the function $\Omega(z;\lambda,\mu)$ has the form:
\begin{align}
	\Omega(z;\lambda,\mu) &= A(z;\lambda,\mu) - \partial_{v}\left[B(z,v;\lambda,\mu) + C(v,z;\lambda,\mu)  \right]_{v=z} \notag \\
	&+ \left[\mathbb{1} \otimes \mathcal{L}(x;\mu), B(z,z;\lambda,\mu)\right] + \left[\mathcal{L}(x;\lambda)\otimes \mathbb{1}, C(z,z;\lambda,\mu)\right]. \label{maillet_Omega_function}
\end{align}		
Thus, the algebra \eqref{maillet_r_s_matrices} is defined now by the two matrices $r(z;\lambda,\mu)$ and $s(z;\lambda,\mu)$, which can now, in general, depend on the coordinates. 

Using these definitions, it is easy to derive the algebra for the transition matrices $T(x,x';\lambda)$:
\begin{align}
	&\{T(x,y;\lambda) \overset{\otimes}{,} T(x',y';\mu) \} \notag \\
&= T(x,x_{0};\lambda) \otimes T(x',x_{0};\mu)\; r(x_{0};\lambda,\mu) \;T(x_{0},y;\lambda) \otimes T(x_{0},y';\mu) \notag \\
&- T(x,y_{0};\lambda) \otimes T(x',y_{0};\mu)\; r(y_{0};\lambda,\mu)\;T(y_{0},y;\lambda) \otimes T(y_{0},y';\mu)  \notag \\
&+ \epsilon(x-x')T(x,x_{0};\lambda) \otimes T(x',x_{0};\mu)\;s(x_{0};\lambda,\mu)\;T(x_{0},y;\lambda) \otimes T(x_{0},y';\mu) \notag \\
&- \epsilon(y'-y)T(x,y_{0};\lambda) \otimes T(x',y_{0};\mu)\;s(y_{0};\lambda,\mu)\;T(y_{0},y;\lambda) \otimes T(y_{0},y';\mu), \label{maillet_T_algebra_original}
\end{align}
where  all the $x,x',y,y'$ are distinct, $x_{0} \equiv min(x,x');\,\, y_{0} \equiv max(y,y')$, and the ordering is chosen such that $x,x' > y,y'$.
The crucial point in this formula is the appearance of the functions $\epsilon(x-x')$ and $\epsilon(y-y')$, which make the algebra \eqref{maillet_T_algebra_original} ill-defined. Indeed, the discontinuity in the algebra \eqref{maillet_T_algebra_original} at coinciding points $x=x'$ (or $y=y'$)  is proportional to the value of the $s$-matrix at this point. Furthermore, as shown in \cite{Maillet:1985ek}, it is not possible to define the Poisson brackets at the coinciding points, such that the Jacobi identity is satisfied. However, it is possible to define \emph{weak} Poisson brackets, via the Maillet's symmetric limit procedure, such that the Jacobi identity, as well as all the standard properties hold. 

The symmetrization procedure and the \emph{weak} Poisson brackets are defined for each $n$-nested brackets of the type:
\begin{align}
\Delta^{n}(x_{i},y_{i};\lambda_{i}) := \left\{ T(x_{1},y_{1};\lambda_{1}) \overset{\otimes}{,} \left\{ \ldots\overset{\otimes}{,}\left\{ T(x_{n},y_{n};\lambda_{n}) \overset{\otimes}{,} \;T(x_{n+1},y_{n+1};\lambda_{n}) \right\} \ldots \right\} \right\}. \label{maillet_nested_brackets}
\end{align}
This quantity is only well-defined if all the points $x_{i}$ and $y_{i}$ are distinct. For coinciding points, one introduces the \emph{weak} Maillet brackets by a point-splitting and symmetrization procedure. For example, for $x_{i}=x$, one defines:
\begin{align}
\Delta^{n}(x,y_{i};\lambda_{i}) := \lim_{\epsilon \rightarrow 0} \frac{1}{(n+1)!} \sum_{\sigma  \,  {\scriptscriptstyle\in} \, \mathbb{P}} \Delta^{n} \left(x + \epsilon \sigma(1),\ldots,x + \epsilon \sigma(n+1),y_{i};\lambda_{i} \right), \label{maillet_weak_brackets}
\end{align}
where $\mathbb{P}$ stands for all possible permutations of $(1,\ldots,n+1)$.
In particular, for  $n=2$, this definition leads to the Maillet brackets between the transition matrices:
\begin{align}
	\{&T(x,y;\lambda) \overset{\otimes}{,} \;T(x,y';\mu) \}_{{M}} \nonumber\\ 
	& := \frac{1}{2} \lim_{\epsilon \rightarrow 0}  \left(\{T(x - \epsilon,y;\lambda) \overset{\otimes}{,} \;T(x + \epsilon,y';\mu) \} + \{T(x + \epsilon,y;\lambda) \overset{\otimes}{,} \;T(x - \epsilon,y';\mu) \} \right).
\end{align}
With such well-defined \emph{weak} brackets, one can show that the algebras for the transition matrices for the equal and adjacent intervals have the form:
\begin{align}
		\{T(x,y;\lambda) \overset{\otimes}{,} \; T(x,y;\mu) \}_{{M}} 
		&= r(x;\lambda,\mu) \; T(x,y;\lambda) \otimes T(x,y;\mu) \label{maillet_equal_intervals} \\
		&- T(x,y;\lambda) \otimes T(x,y;\mu) \;r(y;\lambda,\mu), \notag \\
		\{T(x,y;\lambda) \overset{\otimes}{,} \;T(y,z;\mu) \}_{{M}} &= (T(x,y;\lambda) \otimes \mathbb{1}) \; s(y;\lambda,\mu) \;(\mathbb{1} \otimes T(y,z;\lambda)). \label{maillet_adjacent_intervals}
\end{align}
We refer the reader to the original papers \cite{Maillet:1985ek,Maillet:1985fn,Maillet:1985ec,deVega:1983gy} for detailed discussion and several impressive examples, which can be solved with this prescription. Moreover, it is possible to formulate these relations on the lattice, which makes it possible, in principle, to proceed with quantization \cite{Freidel:1991jv}. 

\subsection{Operator regularization method and the classical limit}
Before considering the generalization of these ideas for the $AAF$ model, we address the validity of using such weakly defined Poisson brackets in the classical theory. Indeed, the construction of the Maillet brackets via the symmetric limit  prescription is not intuitively plausible, and seems somewhat artificial. We claim, however, that Maillet's prescription is the natural consequence of the regularized quantum theory, using the split-point regularization introduced in \cite{Melikyan:2008ab,Melikyan:2010fr}. Although it was demonstrated on the particular example of the Landau-Lifshitz model, the method itself is general enough to be applicable to other singular systems, and, therefore, we consider here only the general ideas, and leave the complete proof for the future publication \cite{melikyan:2013toappear}. 

The  central concept in the operator regularization method is the introduction of the regularized operators $S^{a}_{\mathcal{F}} (x)$:
\begin{equation}
S^{a}_{\mathcal{F}} (x) := \int d\xi\: \mathcal{F} _{\epsilon}(x,\xi) S^{a}(\xi), \label{reg_operators}
\end{equation}
where  $\mathcal{F}_{\epsilon}(x,\xi)$ is some symmetric and smooth function, which depends on some parameter $\epsilon$. The function $\mathcal{F}_{\epsilon}(x,\xi)$ should also satisfy some additional conditions depending on the algebraic structure of the model, such that the singular expressions (for example, the Yang-Baxter equation) due to the operators product at the same point, become well-defined. Moreover, in the $\epsilon \rightarrow 0$ one should restore the original expressions. The theory can then be reformulated in terms of the regularized operators $S^{a}_{\mathcal{F}} (x)$, and only in the end one should remove the regularization $\epsilon \rightarrow 0$. This program has been completed for the Landau-Lifshitz model, and it was shown that the spectrum and the higher order charges can be obtained by using this method \cite{Melikyan:2008ab,Melikyan:2010fr}. In addition, all the quantum charges were shown to be self-adjoint, and the necessary self-adjoint extensions were constructed. We also established a connection between the self-adointness of the quantum charges and the $S$-matrix factorization. It is clear, that the function $\mathcal{F}_{\epsilon}(x,\xi)$  in \eqref{reg_operators} is essentially equivalent to the split-point procedure in the quantum theory, so that the algebra, and other constraints become well-defined. One should consider the quantum theory as fundamental, and the classical theory should follow from the quasi-classical limit: $\hbar \rightarrow 0$. Therefore, such split-point operator regularization in the quantum theory, which makes all the singular expressions well defined, should also appear in the quasi-classical limit. The Maillet symmetric limit prescription \eqref{maillet_weak_brackets} is exactly such split-point regularization, and we conclude that the Maillet's definition of the Poisson brackets is simply the consequence of the regularized operators of the quantum theory. The full proof and details of this argument will be presented in a future publication.

\subsection{Maillet algebra for the $AAF$ model}
We turn now to the algebra for the $L$-operators for the $AAF$ model. As we have seen in the previous section, the algebra has a more general form, and contains terms up to the second derivative of the delta function. Namely, the algebra has the following form:
\begin{align}
	\{ \mathcal{L}(z;\lambda) \overset{\otimes}{,} \mathcal{L}(z';\mu) \} = A(z,z';\lambda) \delta(z-z') +  \displaystyle \sum_{i,j=0}^{2}B_{ij}(z,z';\lambda) \partial_{z}^{i} \partial_{z'}^{j}\delta(z-z'). \label{maillet_aaf_general}
\end{align}
Although we have found the $A$ and $B_{ij}$ functions, we will not present here their explicit expressions, due to their lengthy form. Instead, we will focus on the some of the preliminary essential consequences of the algebraic structure \eqref{maillet_aaf_general} for the bosonic case with arbitrary functions $A$ and $B_{ij}$, leaving the full details and explicit expressions to a separate publication. 

One can recast the algebra \eqref{maillet_aaf_general} in a more convenient form, similar to the one in \eqref{maillet_r_s_matrices}: 
\begin{align}
			\{ \mathcal{L}(z;\lambda) \overset{\otimes}{,} \mathcal{L}(z';\mu) \} &= \left( \partial_{z}r(z;\lambda,\mu) + \left[r(z;\lambda,\mu),\mathcal{L}(z;\lambda) \otimes \ \mathbb{1} + \mathbb{1} \otimes \mathcal{L}(z;\mu) \right] \right. \notag \\
			&\left. +  \left[s_{1}(z;\lambda,\mu),\mathbb{1} \otimes \mathcal{L}(z;\mu) - \mathcal{L}(z;\lambda) \otimes \ \mathbb{1} \right] \right. \notag \\
&\left. + \left[\partial_{z}s_{2}(z;\lambda,\mu), \mathcal{L}(z;\lambda) \otimes \ \mathbb{1} + \mathbb{1} \otimes \mathcal{L}(z;\mu) \right] \right. \notag \\ 
&\left. +\left[ \left[s_{2}(z;\lambda,\mu),\mathcal{L}(z;\lambda) \otimes \ \mathbb{1} \right], \mathbb{1} \otimes \mathcal{L}(z;\mu)\right] \right. \notag \\ 
&\left. +\left[ \left[s_{2}(z;\lambda,\mu), \mathbb{1} \otimes \mathcal{L}(z;\mu)\right], \mathcal{L}(z;\lambda) \otimes \ \mathbb{1}\right] \right) \delta(z-z') \notag  \\
&+s_{1}(z;\lambda,\mu)\partial_{z'}\delta(z-z') + s_{1}(z';\lambda,\mu)\partial_{z}\delta(z-z') \notag  \\
			&+s_{2}(z;\lambda,\mu)\partial_{z'}^{2}\delta(z-z') + s_{2}(z';\lambda,\mu)\partial_{z}^{2}\delta(z-z'), \label{maillet_r_s1_s2_algebra}
	\end{align}
where the $r,s_{1}$ and $s_{2}$ matrices are defined using the functions $A, B_{ij}$, similar to \eqref{maillet_s_matrix}-\eqref{maillet_Omega_function}. The essential difference is the introduction of the additional matrix $s_{2}$, due to the second derivative term in the algebra \eqref{maillet_aaf_general}. 
It is easy to show that the algebra for the transition matrices, that follows from \eqref{maillet_r_s1_s2_algebra} has the form:
\begin{align}
	&\{T(x,y;\lambda) \overset{\otimes}{,} T(x',y';\mu) \} \notag \\
&= T(x,x_{0};\lambda) \otimes T(x',x_{0};\mu)\;u(x_{0};\lambda,\mu)\;T(x_{0},y;\lambda) \otimes T(x_{0},y';\mu) \notag \\
&- T(x,y_{0};\lambda) \otimes T(x',y_{0};\mu)\;u(y_{0};\lambda,\mu)\;T(y_{0},y;\lambda) \otimes T(y_{0},y';\mu)  \notag \\
&+ \epsilon(x-x')T(x,x_{0};\lambda) \otimes T(x',x_{0};\mu)\;v(x_{0};\lambda,\mu)\;T(x_{0},y;\lambda) \otimes T(x_{0},y';\mu) \notag \\
&- \epsilon(y'-y)T(x,y_{0};\lambda) \otimes T(x',y_{0};\mu)\;v(y_{0};\lambda,\mu)\;T(y_{0},y;\lambda) \otimes T(y_{0},y';\mu),
\label{maillet_T_algebra_generalized}
\end{align}
where we have defined:
\begin{align}
u(x;\lambda,\mu) &= r(x;\lambda,\mu) + \partial_{x}s_{2}(x;\lambda,\mu) + \left[ s_{2}(x;\lambda,\mu), \mathcal{L}(x;\lambda) \otimes \mathbb{1} + \mathbb{1} \otimes \mathcal{L}(x;\mu)   \right] 
\label{maillet_u_matrix} \\
v(x;\lambda,\mu) &= s_{1}(x;\lambda,\mu) + \left[ s_{2}(x;\lambda,\mu), \mathcal{L}(x;\lambda) \otimes \mathbb{1} - \mathbb{1} \otimes \mathcal{L}(x;\mu)   \right] \label{maillet_v_matrix}
\end{align}

This algebra has exactly the same form as the algebra \eqref{maillet_T_algebra_original}, where the $r(x;\lambda,\mu)$ and  $s(x;\lambda,\mu)$ functions are replaced by the $u(x;\lambda,\mu)$ and $v(x;\lambda,\mu)$ functions. Hence, one can immediately write the algebra for transition matrices for the equal and adjacent intervals from \eqref{maillet_T_algebra_generalized} by replacing $r \rightarrow u$ and $ s \rightarrow v$ in \eqref{maillet_equal_intervals} and \eqref{maillet_adjacent_intervals}, as well as use the general results of the previous sections, including the construction of the symmetric limit prescription for nested brackets. Thus, we conclude that even though the algebra for $L$-operators \eqref{maillet_r_s1_s2_algebra} is modified by the additional terms depending on the $s_2(x;\lambda,\mu)$ matrix, the resulting algebra for the transition matrices \eqref{maillet_T_algebra_generalized} is the same, and the complication introduced by the second derivative term is only technical. The essential steps of the Maillet prescription can be repeated without any conceptual complications for the graded algebra as well, which is the case of the $AAF$ model, the generalization is straightforward. Finally, we note that the lattice version of the algebra \eqref{maillet_T_algebra_generalized} is  known as well \cite{Freidel:1991jv}, and may in principle be used to construct the lattice version of the $AAF$ model. The full details of this analysis will be presented in the future publication \cite{melikyan:2013toappear}.

\section{Conclusion}\label{conclusion}
We have considered in this paper the classical integrability of the $AAF$ model, starting from the first principles of the inverse scattering method. The necessity of this consideration was stimulated by the fact that the perturbative calculation still do not provide the complete information about the integrable properties of the $AAF$ model. For example, the perturbative approach was not sensitive to the rather complicated structure of the Dirac brackets, which, therefore, was not taken into account. In the process of our consideration, we have proved that the Lax operator surprisingly admits a simpler $2 \times 2$ representation. This is especially curious, since the much simpler Thirring model has a more complex $3 \times 3$ known representation. We have also found that the constraint on the coupling constants $g_{2}^{2} = g_{3}$, which was derived in \cite{Melikyan:2011uf} from the $S$-matrix factorization condition, must also be satisfied in the classical theory, and that for no extension of the Lax connections integrability holds without this constraint.

The algebraic structure of the $AAF$ model, however, is rather complicated and has a highly non-ultralocal form, which  contains terms up to the second derivative in delta function in the algebra of the Lax operators. This led us to the generalization of the regularization technique due to Maillet. Here, it was necessary to introduce three matrices $r, s_{1}$ and $s_{2}$ to properly encode the algebraic structure. We have also derived the algebra for transition matrices for the $AAF$ model, and have shown it to have exactly the same form as it is the case of the usual non-ultralocality, with modified $r$- and $s$-matrices.  Strictly speaking, although we have obtained all the general expressions, further progress depends on the analysis of the equation defining the $r$-matrix, and finding its local solutions. This will allow one to proceed in the manner similar to, for example, the complex sine-Gordon model, where the $r$-matrix has indeed a local, coordinate independent form, and the model can be solved exactly in the infinite space limit \cite{Maillet:1985ek}. 

We also obtained in the process the fermionic counterpart of the Wadati model, which introduces an interesting toy-model for testing the generalized Maillet algebra, as it seems possible to explicitly compute some relevant quantities, such as the transition matrix and its infinite line limit.

We have also proposed that the symmetric limit prescription of Maillet has its origin in the regularized quantum theory. Indeed, based on the method proposed in \cite{Melikyan:2010fr}, the quantum theory is made well-defined via operator regularization. Maillet algebra is then obtained in the quasi-classical limit. This is an interesting direction which will be explored in more details in the future.

The problem of obtaining an alternative Lax connection, for which the algebra becomes ultralocal is open. In principle there may exist such a connection, which does not follow from the reduction of the strings on $AdS_5 \times S^5$ to the $\mathfrak{su}(1|1)$ subsector. It would also be interesting to obtain the equivalent model via bosonization procedure. Indeed, it is easy to see, by examining the constraints of the gauge fixing procedure in \cite{Alday:2005jm}, that the additional higher order non-ultralocality appears in the algebra due to eliminating  the bosonic fields  in favor of the fermionic ones. Thus, it is very desirable to obtain the bosonic version of the $AAF$  model, which should be less non-ultralocal, and may make the integrability analysis a much simpler task. 

\section*{Appendices} \addcontentsline{toc}{section}{Appendices}
\appendix 
\section{Notations}\label{app_notations}
We use the following representation for the two-dimensional Dirac matrices: 
\begin{equation}
	\label{dirac_matrices} \rho^0 = \left( 
	\begin{array}{cc}
		-1 & 0 \\
		0 & 1 
	\end{array}
	\right), \quad \rho^1 = \left( 
	\begin{array}{cc}
		0 & i \\
		i & 0 
	\end{array}
	\right)\quad \text{and} \quad \rho^5 = \rho^0 \rho^1.
\end{equation}

Following the notations in \cite{Alday:2005jm}, we choose here the sign of $\kappa$ as:
\begin{equation}
	 \kappa = \frac{\sqrt{\lambda}}{2} \label{kappa_choice}.
\end{equation}
In this case the functions $l_{i}$ have the following dependence on the spectral parameter $\mu$ (see the discussion in \cite{Alday:2005jm,Alday:2005gi}):
\begin{align}
	l_{0} = 1,\quad l_{1} = \frac{1+\mu^{2}}{1-\mu^2},\quad l_{2} = s_{1}\frac{2\mu}{1-\mu^2}, \quad  l_{3} = s_{2}\frac{1}{\sqrt{1-\mu^2}} \quad \text{and} \quad l_{4} = s_{3}\frac{\mu}{\sqrt{1-\mu^2}}, \label{l_functions}
\end{align}
with the following choice of parameters:
\begin{align}
    s_{1}+s_{2}s_{3} = 0 \quad \text{and} \quad (s_{2})^{2} = (s_{3})^{2} = 1. 
\end{align}
Alternatively, one could use the parametrization:
\begin{align}
	l_{0} = 1,\quad l_{1} = \cosh{(2\beta)},\quad l_{2} = -\sinh{(2\beta)},\quad l_{3}=\cosh{(\beta)} \quad \text{and} \quad l_{4}=\sinh{(\beta)}. \label{alternative_l_functions}
\end{align}
\section{Equations of Motion in the reduced form}\label{app_eoms_reduced}
In this appendix we give the equations of motion for the AAF model in the reduced form. As explained in the main text, they are derived from the original equations of motion \eqref{eom_psi1} and \eqref{eom_psi1} by recursively eliminating the time derivatives present in the cubic and higher order terms, through multiple usage of the equations of motion.

The reduced equation for  $\psi_{1}$, following from the $AAF$ Lagrangian, has the form:
\begin{align}
&iJ\dot{\psi_{1}} - \sqrt{\lambda} \psi_{2}' + J\psi_{1} -\frac{\sqrt{\lambda}g_{2}}{2J}\left[ \sqrt{\lambda} \left(2\psi_{1}' \psi_{2}^{*} \psi_{2}' - \psi_{1}'^{*}\psi_{2}\psi_{2}' -\psi_{1}'^{*}\psi_{1}\psi_{1}' - \psi_{2}'^{*}\psi_{2}\psi_{1}' + \psi_{2}'^{*}\psi_{1}\psi_{2}' \right)\right. \notag \\
&\left. +J\left( \psi_{1}\psi_{2}^{*}\psi_{1}' - \psi_{2}\psi_{2}^{*}\psi_{2}' +\psi_{1}^{*}\psi_{2}\psi_{1}' - \psi_{1}^{*}\psi_{1}\psi_{2}' \right) \right] - \frac{ \lambda(g_{2})^{2}}{4J^{2}}\left[\sqrt{\lambda} \left( 3 \psi_{1}'^{*}\psi_{2}\psi_{2}' \psi_{2}^{*} \psi_{1}'  \right. \right. \notag \\
&\left. \left. - 3\psi_{2}'^{*}\psi_{1}\psi_{2}' \psi_{2}^{*} \psi_{1}' -2\psi_{2}'^{*}\psi_{1}'^{*}\psi_{2} \psi_{1}\psi_{2}'- \psi_{1}'\psi_{1}'^{*}\psi_{1}\psi_{1}^{*} \psi_{2}' - \psi_{2}'^{*}\psi_{1}^{*}\psi_{2}\psi_{1}'\psi_{2}' \right) + J \left( 3\psi_{1}^{*}\psi_{1}'^{*}\psi_{2}\psi_{1}\psi_{2}' \right. \right. \notag \\
&\left. \left. + 2\psi_{1}^{*}\psi_{1}\psi_{2}'\psi_{2}^{*}\psi_{1}' -\psi_{2}^{*}\psi_{1}'^{*}\psi_{1}\psi_{2}\psi_{1}' +\psi_{1}\psi_{2}'^{*}\psi_{1}^{*}\psi_{2}\psi_{1}' - \psi_{2}^{*}\psi_{2}'^{*}\psi_{2}\psi_{1}\psi_{2}' \right)\right]  \notag \\
&- \frac{\sqrt{\lambda}g_{3}}{8J} \left[ \sqrt{\lambda} \left( 3\psi_{1}'^{*}\psi_{2}^{*}\psi_{1}\psi_{2}\psi_{1}' -3\psi_{2}'^{*}\psi_{2}^{*}\psi_{1}\psi_{2}\psi_{2}' + 2\psi_{1}^{*}\psi_{2}^{*}\psi_{1}\psi_{2}'\psi_{1}' - \psi_{1}'^{*}\psi_{1}^{*}\psi_{1}\psi_{2}\psi_{2}' \right. \right.\notag \\ 
&\left. \left. + \psi_{1}^{*}\psi_{2}'^{*}\psi_{1}\psi_{2}\psi_{1}' \right)+2J\psi_{1}^{*}\psi_{2}^{*}\psi_{1}\psi_{2}\psi_{2}' \right] +\frac{(\lambda)^{3/2}(g_{2})^{3}}{8J^{3}}\left[ -8\sqrt{\lambda}\psi_{2}'^{*}\psi_{1}'^{*}\psi_{2}\psi_{1}\psi_{2}'\psi_{2}^{*}\psi_{1}' \right. \notag \\
& \left. + 4J\psi_{1}^{*}\psi_{1}'^{*}\psi_{2}\psi_{1}\psi_{2}'\psi_{2}^{*}\psi_{1}'   \right] +\frac{(\lambda)^{3/2}g_{2}g_{3}}{4J^{2}}\left[\psi_{1}'\psi_{1}^{*}\psi_{1}'^{*}\psi_{1}\psi_{2}\psi_{2}^{*}\psi_{2}'   \right] = 0, \label{eom_reduced_psi1}
\end{align}
while the reduced equation for $\psi_{2}$, which follows from the $AAF$ Lagrangian, is:
\begin{align}
	&iJ\dot{\psi_{2}} - \sqrt{\lambda} \psi_{1}' + J\psi_{2} -\frac{\sqrt{\lambda}g_{2}}{2J}\left[ \sqrt{\lambda} \left( -2\psi_{2}' \psi_{1}^{*} \psi_{1}' + \psi_{2}'^{*}\psi_{2}\psi_{2}' + \psi_{2}'^{*}\psi_{1}\psi_{1}' - \psi_{1}'^{*}\psi_{2}\psi_{1}' + \psi_{1}'^{*}\psi_{1}\psi_{2}' \right) \right. \notag \\
&\left. + J \left( \psi_{2}^{*}\psi_{2}\psi_{1}' - \psi_{2}^{*}\psi_{2}\psi_{2}' -\psi_{2}\psi_{1}^{*}\psi_{2}' + \psi_{1}\psi_{1}^{*}\psi_{1}'  \right) \right] + \frac{ \lambda(g_{2})^{2}}{4J^{2}}\left[ \sqrt{\lambda} \left( -3 \psi_{2}'^{*}\psi_{1}\psi_{1}^{*} \psi_{2}' \psi_{1}' \right. \right. \notag \\
&\left. \left. + 3 \psi_{1}'^{*}\psi_{1}^{*}\psi_{2}' \psi_{2}\psi_{1}' +2\psi_{1}'^{*}\psi_{2}'^{*}\psi_{1}\psi_{2}\psi_{1}'+ \psi_{2}'^{*}\psi_{2}\psi_{2}^{*}\psi_{1}'\psi_{2}' + \psi_{1}'\psi_{1}'^{*}\psi_{2}^{*}\psi_{1}\psi_{2}' \right) +J \left(- 3\psi_{2}^{*}\psi_{2}'^{*}\psi_{1}\psi_{2}\psi_{1}' \right. \right. \notag \\
&\left. \left. - 2\psi_{2}^{*}\psi_{2}\psi_{1}'\psi_{1}^{*}\psi_{2}' + \psi_{2}'^{*}\psi_{1}\psi_{1}^{*}\psi_{2}'\psi_{2} + \psi_{1}^{*}\psi_{1}'^{*}\psi_{1}\psi_{2}\psi_{1}' - \psi_{2}\psi_{1}'^{*}\psi_{2}^{*}\psi_{1}\psi_{2}' \right)\right]  \notag \\
&- \frac{\sqrt{\lambda}g_{3}}{8J} \left[ \sqrt{\lambda} \left( 3\psi_{1}'^{*}\psi_{1}^{*}\psi_{1}\psi_{2}\psi_{1}' +3\psi_{1}^{*}\psi_{2}'^{*}\psi_{1}\psi_{2}\psi_{2}'  - 2\psi_{1}^{*}\psi_{2}^{*}\psi_{2}\psi_{1}'\psi_{2}' + \psi_{2}'^{*}\psi_{2}^{*}\psi_{1}\psi_{2}\psi_{1}' \right. \right. \notag \\
&\left. \left. - \psi_{2}^{*}\psi_{1}'^{*}\psi_{1}\psi_{2}\psi_{2}'\right) +2J\psi_{1}^{*}\psi_{2}^{*}\psi_{1}\psi_{2}\psi_{1}' \right] +\frac{(\lambda)^{3/2}(g_{2})^{3}}{8J^{3}}\left[ 8\sqrt{\lambda}\psi_{1}'^{*}\psi_{1}^{*}\psi_{2}'\psi_{2}'^{*}\psi_{1}\psi_{2}\psi_{1}' \right. \notag \\ 
&\left. - 4J\psi_{2}^{*}\psi_{1}^{*}\psi_{2}'\psi_{2}'^{*}\psi_{1}\psi_{2}\psi_{1}'   \right] -\frac{(\lambda)^{3/2}g_{2}g_{3}}{4J^{2}}\left[\psi_{2}'\psi_{2}^{*}\psi_{2}'^{*}\psi_{2}\psi_{1}\psi_{1}^{*}\psi_{1}'   \right] =0.\label{eom_reduced_psi2}
\end{align}

The dynamical equation for $\psi_{1}$, following from the zero-curvature condition, has the form:
\begin{align}
	&iJ\dot{\psi_{1}} - \sqrt{\lambda} \psi_{2}' + J \psi_{1} - \frac{i\sqrt{\lambda}}{4}\left[2\psi_{2}^{*}\dot{\psi}_{2}\psi_{2}' - \dot{\psi}_{1}^{*}\psi_{1}\psi_{2}' + 2\psi_{1}'^{*}\dot{\psi}_{1}\psi_{2} - \dot{\psi}_{1}^{*}\psi_{1}'\psi_{2} - \psi_{1}^{*}\psi_{1}'\dot{\psi}_{2} + \psi_{1}'^{*}\psi_{1}\dot{\psi}_{2} \right. \notag \\
&\left. - \psi_{2}^{*}\psi_{1}'\dot{\psi}_{1} + \psi_{2}'^{*}\psi_{1}\dot{\psi}_{1}   \right] + \frac{iJ}{4}\left[2\psi_{2}^{*}\dot{\psi}_{2}\psi_{1} + \psi_{1}^{*}\psi_{1}\dot{\psi}_{1} - \psi_{2}^{*}\psi_{2}\dot{\psi}_{1}\right] - \frac{\lambda}{4J}\left[\psi_{2}^{*}\psi_{2}'\psi_{1}' - \psi_{1}'^{*}\psi_{1}\psi_{1}' - \psi_{2}'^{*}\psi_{2}\psi_{1}'\right] \notag \\
&+\frac{\sqrt{\lambda}}{4}\left[\psi_{2}^{*}\psi_{1}'\psi_{1}
+\psi_{2}^{*}\psi_{2}\psi_{2}' + \psi_{1}'^{*}\psi_{2}\psi_{1}\right]  - \frac{3J}{4}\left[\psi_{2}^{*}\psi_{2}\psi_{1}\right] - \frac{\lambda}{16J}\left[ - 4\psi_{2}^{*}\dot{\psi}_{2}\psi_{1}^{*}\psi_{1}\psi_{2}' - \dot{\psi}_{1}^{*}\psi_{1}\psi_{2}^{*}\psi_{2}\psi_{2}' \right. \notag \\
&\left. + \dot{\psi}_{2}^{*}\psi_{2}\psi_{1}^{*}\psi_{1}\psi_{2}' + \psi_{1}^{*}\psi_{1}'\psi_{2}^{*}\psi_{2}\dot{\psi}_{2} - \psi_{1}'^{*}\psi_{1}\psi_{2}^{*}\psi_{2}\dot{\psi}_{2} - \psi_{2}'^{*}\psi_{2}\psi_{1}^{*}\psi_{1}\dot{\psi}_{2} + 2\psi_{1}^{*}\psi_{1}'\dot{\psi}_{1}^{*}\psi_{1}\psi_{2} - \psi_{1}^{*}\dot{\psi}_{1}\psi_{1}'^{*}\psi_{1}\psi_{2} \right. \notag \\
&\left. +2\psi_{1}^{*}\dot{\psi}_{1}\psi_{2}^{*}\psi_{2}'\psi_{2} + \psi_{2}'^{*}\psi_{1}\psi_{2}^{*}\psi_{2}\dot{\psi}_{1} + \psi_{2}^{*}\psi_{1}'\psi_{1}^{*}\psi_{1}\dot{\psi}_{1}\right] - \frac{\lambda}{16J}\left[-3\psi_{2}^{*}\psi_{2}'\psi_{1}^{*}\psi_{1}\psi_{1}' - \psi_{1}'^{*}\psi_{1}\psi_{2}^{*}\psi_{2}\psi_{1}' \right. \notag \\
&\left. +\psi_{2}'^{*}\psi_{2}\psi_{1}^{*}\psi_{1}\psi_{1}' -2\psi_{2}^{*}\psi_{2}'\psi_{2}'^{*}\psi_{2}\psi_{1}\right] = 0, \label{zcc_eom_psi1}
\end{align}
while the dynamical equation for $\psi_{2}$, which follows from the zero-curvature condition, is:
\begin{align}
	&-iJ\dot{\psi_{2}} - \sqrt{\lambda} \psi_{1}' + J \psi_{2} - \frac{i\sqrt{\lambda}}{4}\left[2\psi_{1}^{*}\dot{\psi}_{1}\psi_{1}' - \dot{\psi}_{2}^{*}\psi_{2}\psi_{1}' + 2\psi_{2}'^{*}\dot{\psi}_{2}\psi_{1} - \dot{\psi}_{2}^{*}\psi_{2}'\psi_{1} - \psi_{2}^{*}\psi_{2}'\dot{\psi}_{1} + \psi_{2}'^{*}\psi_{2}\dot{\psi}_{1} \right. \notag \\
&\left. - \psi_{1}^{*}\psi_{2}'\dot{\psi}_{2} + \psi_{1}'^{*}\psi_{2}\dot{\psi}_{2}   \right] + \frac{iJ}{4}\left[2\psi_{1}^{*}\dot{\psi}_{1}\psi_{2} + \psi_{2}^{*}\psi_{2}\dot{\psi}_{2} - \psi_{1}^{*}\psi_{1}\dot{\psi}_{2}\right] - \frac{\lambda}{4J}\left[\psi_{1}^{*}\psi_{1}'\psi_{2}' - \psi_{2}'^{*}\psi_{2}\psi_{2}' - \psi_{1}'^{*}\psi_{1}\psi_{2}'\right] \notag \\
&-\frac{\sqrt{\lambda}}{4}\left[\psi_{1}^{*}\psi_{2}'\psi_{2}
+\psi_{2}'^{*}\psi_{1}\psi_{2} + \psi_{1}^{*}\psi_{1}\psi_{1}'\right]  + \frac{3J}{4}\left[\psi_{1}^{*}\psi_{1}\psi_{2}\right] - \frac{\lambda}{16J}\left[4\psi_{1}^{*}\dot{\psi}_{1}\psi_{2}^{*}\psi_{2}\psi_{1}' + 2\psi_{2}^{*}\dot{\psi}_{2}\psi_{1}^{*}\psi_{1}\psi_{1}' \right. \notag \\
&\left. -\dot{\psi}_{1}^{*}\psi_{1}\psi_{2}^{*}\psi_{2}\psi_{1}' +\dot{\psi}_{2}^{*}\psi_{2}\psi_{1}^{*}\psi_{1}\psi_{1}' + \psi_{1}^{*}\dot{\psi}_{1}\psi_{2}^{*}\psi_{2}'\psi_{1} + \psi_{2}'^{*}\psi_{1}\psi_{1}^{*}\dot{\psi}_{1}\psi_{2} + \psi_{1}'^{*}\psi_{1}\psi_{2}^{*}\psi_{2}\dot{\psi}_{1} -2\psi_{2}^{*}\psi_{2}'\dot{\psi}_{2}^{*}\psi_{2}\psi_{1} \right. \notag \\
&\left. +\psi_{2}'^{*}\psi_{2}\psi_{2}^{*}\dot{\psi}_{2}\psi_{1} - \psi_{1}^{*}\psi_{2}'\psi_{2}^{*}\psi_{2}\dot{\psi}_{2} - \psi_{1}'^{*}\psi_{2}\psi_{1}^{*}\psi_{1}\dot{\psi}_{2} \right] - \frac{\lambda}{16J}\left[-3\psi_{1}^{*}\psi_{1}'\psi_{2}^{*}\psi_{2}\psi_{2}' - \psi_{2}'^{*}\psi_{2}\psi_{1}^{*}\psi_{1}\psi_{2}' \right. \notag \\
&\left. +\psi_{1}'^{*}\psi_{1}\psi_{2}^{*}\psi_{2}\psi_{2}' -2\psi_{1}^{*}\psi_{1}'\psi_{1}'^{*}\psi_{1}\psi_{2}\right] = 0.
\label{zcc_eom_psi2}
\end{align}

\section{The list of $M^{(ij)}$ functions}\label{app_M_functions}
Here we give the explicit form of the $M_{3}^{(ij)}$, $M_{4}^{(ij)}$, $M_{13}^{(ij)}$, $M_{14}^{(ij)}$, $M_{23}^{(ij)}$ and $M_{24}^{(ij)}$ functions (see \eqref{M_functions}).
\begin{itemize}
\item $(i=1;j=1)$:
\begin{align}
	M_{3}^{(11)} &= -2i \partial_{0}(\alpha_{0}\gamma_{\sigma}\chi_{21}) + 2i\partial_{1}(\alpha_{0}\gamma_{\tau}\chi_{21}), \label{M_3_11}\\
	M_{4}^{(11)} &= 2\partial_{0}(\alpha_{0}\gamma_{\sigma}\tilde{\chi}_{21}) - 2\partial_{1}(\alpha_{0}\gamma_{\tau}\tilde{\chi}_{21}), \label{M_4_11}\\
    M_{13}^{(11)} &= -2i \gamma_{\tau}\partial_{1}(\alpha_{0}\chi_{21}) + 2i\gamma_{\sigma}\partial_{0}(\alpha_{0}\chi_{21}), \label{M_13_11}\\
	M_{14}^{(11)} &= -4 \gamma_{\tau}\partial_{1}(\alpha_{0}\tilde{\chi}_{21}) + 4\gamma_{\sigma}\partial_{0}(\alpha_{0}\tilde{\chi}_{21}), \label{M_14_11}\\
	M_{23}^{(11)} &= -\frac{4\sqrt{\lambda}}{J}\gamma_{\sigma}\alpha_{0}\gamma_{\sigma}\chi_{11} - \frac{2i\sqrt{\lambda}}{J}\gamma_{\sigma}\partial_{1}(\alpha_{0}\chi_{21}) + \frac{i}{\sqrt{\lambda}}\zeta\alpha_{0}\gamma_{\tau}\chi_{11} - \frac{1}{2\sqrt{\lambda}}\zeta\partial_{0}(\alpha_{0}\chi_{21}), \label{M_23_11}\\
	M_{24}^{(11)} &= -\frac{4i\sqrt{\lambda}}{J}\gamma_{\sigma}\alpha_{0}\gamma_{\sigma}\tilde{\chi}_{11} - \frac{2\sqrt{\lambda}}{J}\gamma_{\sigma}\partial_{1}(\alpha_{0}\tilde{\chi}_{21}) - \frac{i}{\sqrt{\lambda}}\zeta\alpha_{0}\gamma_{\tau}\tilde{\chi_{11}} + \frac{i}{2\sqrt{\lambda}}\zeta\partial_{0}(\alpha_{0}\tilde{\chi}_{21}).  \label{M_24_11}
\end{align}
\item $(i=1;j=2)$:
\begin{align}
	M_{3}^{(12)} &= -2i \partial_{0}(\alpha_{0}\gamma_{\sigma}\chi_{22}) + 2i\partial_{1}(\alpha_{0}\gamma_{\tau}\chi_{22}), \label{M_3_12}\\
	M_{4}^{(12)} &= 2\partial_{0}(\alpha_{0}\gamma_{\sigma}\tilde{\chi}_{22}) - 2\partial_{1}(\alpha_{0}\gamma_{\tau}\tilde{\chi}_{22}), \label{M_4_12}\\
    M_{13}^{(12)} &= -2i \gamma_{\tau}\partial_{1}(\alpha_{0}\chi_{22}) + 2i\gamma_{\sigma}\partial_{0}(\alpha_{0}\chi_{22}), \label{M_13_12}\\
	M_{14}^{(12)} &= -4 \gamma_{\tau}\partial_{1}(\alpha_{0}\tilde{\chi}_{22}) + 4\gamma_{\sigma}\partial_{0}(\alpha_{0}\tilde{\chi}_{22}), \label{M_14_12}\\
	M_{23}^{(12)} &= -\frac{4\sqrt{\lambda}}{J}\gamma_{\sigma}\alpha_{0}\gamma_{\sigma}\chi_{12} - \frac{2i\sqrt{\lambda}}{J}\gamma_{\sigma}\partial_{1}(\alpha_{0}\chi_{22}) + \frac{i}{\sqrt{\lambda}}\zeta\alpha_{0}\gamma_{\tau}\chi_{12} - \frac{1}{2\sqrt{\lambda}}\zeta\partial_{0}(\alpha_{0}\chi_{22}), \label{M_23_12}\\
	M_{24}^{(11)} &= -\frac{4i\sqrt{\lambda}}{J}\gamma_{\sigma}\alpha_{0}\gamma_{\sigma}\tilde{\chi}_{12} - \frac{2\sqrt{\lambda}}{J}\gamma_{\sigma}\partial_{1}(\alpha_{0}\tilde{\chi}_{22}) - \frac{i}{\sqrt{\lambda}}\zeta\alpha_{0}\gamma_{\tau}\tilde{\chi_{12}} + \frac{i}{2\sqrt{\lambda}}\zeta\partial_{0}(\alpha_{0}\tilde{\chi}_{22}).  \label{M_24_12}
\end{align}
\item $(i=2;j=1)$:
\begin{align}
	M_{3}^{(21)} &= 2i \partial_{0}(\alpha_{0}\gamma_{\sigma}\chi_{11}) - 2i\partial_{1}(\alpha_{0}\gamma_{\tau}\chi_{11}), \label{M_3_21}\\
	M_{4}^{(21)} &= -2\partial_{0}(\alpha_{0}\gamma_{\sigma}\tilde{\chi}_{11}) + 2\partial_{1}(\alpha_{0}\gamma_{\tau}\tilde{\chi}_{11}), \label{M_4_21}\\
    M_{13}^{(21)} &= 2i \gamma_{\tau}\partial_{1}(\alpha_{0}\chi_{11}) - 2i\gamma_{\sigma}\partial_{0}(\alpha_{0}\chi_{11}), \label{M_13_21}\\
	M_{14}^{(21)} &= 4 \gamma_{\tau}\partial_{1}(\alpha_{0}\tilde{\chi}_{11}) - 4\gamma_{\sigma}\partial_{0}(\alpha_{0}\tilde{\chi}_{11}), \label{M_14_21}\\
	M_{23}^{(21)} &= -\frac{4\sqrt{\lambda}}{J}\gamma_{\sigma}\alpha_{0}\gamma_{\sigma}\chi_{21} + \frac{2i\sqrt{\lambda}}{J}\gamma_{\sigma}\partial_{1}(\alpha_{0}\chi_{11}) + \frac{i}{\sqrt{\lambda}}\zeta\alpha_{0}\gamma_{\tau}\chi_{21} + \frac{1}{2\sqrt{\lambda}}\zeta\partial_{0}(\alpha_{0}\chi_{11}), \label{M_23_21}\\
	M_{24}^{(21)} &= -\frac{4i\sqrt{\lambda}}{J}\gamma_{\sigma}\alpha_{0}\gamma_{\sigma}\tilde{\chi}_{21} + \frac{2\sqrt{\lambda}}{J}\gamma_{\sigma}\partial_{1}(\alpha_{0}\tilde{\chi}_{11}) - \frac{i}{\sqrt{\lambda}}\zeta\alpha_{0}\gamma_{\tau}\tilde{\chi_{21}} - \frac{i}{2\sqrt{\lambda}}\zeta\partial_{0}(\alpha_{0}\tilde{\chi}_{11}).  \label{M_24_21}
\end{align}
\item $(i=2;j=2)$:
\begin{align}
	M_{3}^{(22)} &= 2i \partial_{0}(\alpha_{0}\gamma_{\sigma}\chi_{12}) - 2i\partial_{1}(\alpha_{0}\gamma_{\tau}\chi_{12}), \label{M_3_22}\\
	M_{4}^{(22)} &= -2\partial_{0}(\alpha_{0}\gamma_{\sigma}\tilde{\chi}_{12}) + 2\partial_{1}(\alpha_{0}\gamma_{\tau}\tilde{\chi}_{12}), \label{M_4_22}\\
    M_{13}^{(22)} &= 2i \gamma_{\tau}\partial_{1}(\alpha_{0}\chi_{12}) - 2i\gamma_{\sigma}\partial_{0}(\alpha_{0}\chi_{12}), \label{M_13_22}\\
	M_{14}^{(22)} &= 4 \gamma_{\tau}\partial_{1}(\alpha_{0}\tilde{\chi}_{12}) - 4\gamma_{\sigma}\partial_{0}(\alpha_{0}\tilde{\chi}_{12}), \label{M_14_22}\\
	M_{23}^{(22)} &= -\frac{4\sqrt{\lambda}}{J}\gamma_{\sigma}\alpha_{0}\gamma_{\sigma}\chi_{22} + \frac{2i\sqrt{\lambda}}{J}\gamma_{\sigma}\partial_{1}(\alpha_{0}\chi_{12}) + \frac{i}{\sqrt{\lambda}}\zeta\alpha_{0}\gamma_{\tau}\chi_{22} + \frac{1}{2\sqrt{\lambda}}\zeta\partial_{0}(\alpha_{0}\chi_{12}),\label{M_23_22}\\
	M_{24}^{(22)} &= -\frac{4i\sqrt{\lambda}}{J}\gamma_{\sigma}\alpha_{0}\gamma_{\sigma}\tilde{\chi}_{22} + \frac{2\sqrt{\lambda}}{J}\gamma_{\sigma}\partial_{1}(\alpha_{0}\tilde{\chi}_{12}) - \frac{i}{\sqrt{\lambda}}\zeta\alpha_{0}\gamma_{\tau}\tilde{\chi_{22}} - \frac{i}{2\sqrt{\lambda}}\zeta\partial_{0}(\alpha_{0}\tilde{\chi}_{12}).  \label{M_24_22}
\end{align}
\end{itemize}

With the choice of the $\nu_{i},\tilde{\nu_{i}}$ fields \eqref{nu_aaf_choice}, it is easy to show the following equivalence relations between the equations \eqref{M_3_13} - \eqref{M_4_14}:
\begin{align}
	M_{3}^{(11)} + M_{13}^{(11)} = 0 \quad &\Longleftrightarrow \quad	M_{4}^{(21)} - M_{14}^{(21)} = 0, \notag \\
	M_{3}^{(12)} + M_{13}^{(12)} = 0 \quad &\Longleftrightarrow \quad	M_{4}^{(22)} - M_{14}^{(22)} = 0, \notag \\
	M_{3}^{(21)} + M_{13}^{(21)} = 0 \quad &\Longleftrightarrow \quad	M_{4}^{(11)} - M_{14}^{(11)} = 0, \notag \\
	M_{3}^{(22)} + M_{13}^{(22)} = 0 \quad &\Longleftrightarrow \quad	M_{4}^{(12)} - M_{14}^{(12)} = 0, \notag \\
	M_{13}^{(11)} - M_{24}^{(11)} = 0 \quad &\Longleftrightarrow \quad	M_{14}^{(21)} - M_{23}^{(21)} = 0, \notag \\
	M_{13}^{(12)} - M_{24}^{(12)} = 0 \quad &\Longleftrightarrow \quad	M_{14}^{(22)} - M_{23}^{(22)} = 0, \notag \\
	M_{13}^{(21)} - M_{24}^{(21)} = 0 \quad &\Longleftrightarrow \quad	M_{14}^{(11)} - M_{23}^{(11)} = 0, \notag \\
	M_{13}^{(22)} - M_{24}^{(22)} = 0 \quad &\Longleftrightarrow \quad	M_{14}^{(12)} - M_{23}^{(12)} = 0. \notag
\end{align} 
These relations show that the independent equations are the ones gives in the formulas \eqref{M_3_13a} and \eqref{M_13_24a}.

\section{Useful identities}\label{app_identities}

We collect here some useful identities used throughout the text. These identities can be proved by direct verification, with the use of the equations of motion in the form  \eqref{eom_reduced_psi1} and \eqref{eom_reduced_psi2}. 

The following identity is equivalent to the constraint \eqref{M3_M13_constraint}
\begin{equation}
	\psi_{1}'^{*} \dot{\psi}_{1} - 	\dot{\psi}_{1}^{*}\psi_{1}' + 	\psi_{2}'^{*} \dot{\psi}_{2} - 	\dot{\psi}_{2}^{*}\psi_{2}' = -i\partial_{1}\left(\bar{\psi} \psi\right). \label{identity_1}
\end{equation}
The identity \eqref{B_constraint} can be derived by utilizing the following identities:
\begin{align}
	\psi_{1} \dot{\psi}_{2}^{*} + \psi_{2}^{*}\dot{\psi}_{1} - \psi_{1}^{*}\dot{\psi}_{2} -\psi_{2}\dot{\psi}_{1}^{*} &= -\frac{i\sqrt{\lambda}}{J}\left( \psi_{2}^{*}\psi_{2}' - \psi_{2}'^{*}\psi_{2} + \psi_{1}^{*}\psi_{1}' - \psi_{1}'^{*}\psi_{1} \right), \label{identity_2} \\
	\dot{\psi_{2}}^{*}\psi_{1}' - \psi_{2}'^{*}\dot{\psi}_{1} - \dot{\psi_{1}}^{*}\psi_{2}' + \psi_{1}'^{*}\dot{\psi}_{2} &= - \frac{J}{\sqrt{\lambda}} \left(\dot{\psi}_{1}^{*}\psi_{1} + \psi_{1}^{*}\dot{\psi}_{1} - \dot{\psi}_{2}^{*}\psi_{2} - \psi_{2}^{*}\dot{\psi}_{2} \right). \label{identity_3}
\end{align}
The constraint \eqref{A_constraint} is obtained by computing explicitly the left and the right hand sides, and showing that both are equal to:
\begin{align}
	&\frac{i\lambda}{J}\left[ 
\psi_{1}^{*}\psi_{2}'\psi_{2}\psi_{2}'^{*} + \psi_{2}'\psi_{2}^{*}\psi_{2}'^{*}\psi_{1} - \psi_{1}'\psi_{1}'^{*}\psi_{1}\psi_{2}^{*} -
\psi_{2}\psi_{1}^{*}\psi_{1}'\psi_{1}'^{*} +
2\psi_{1}'\psi_{2}'^{*}\psi_{2}^{*}\psi_{2} +
2\psi_{1}'^{*}\psi_{2}\psi_{2}^{*}\psi_{2}' \right. \notag \\
&\left. + 2\psi_{1}'^{*}\psi_{1}^{*}\psi_{1}\psi_{2}' +
2\psi_{2}'^{*}\psi_{1}^{*}\psi_{1}\psi_{1}' \right] - \frac{2i\sqrt{\lambda}}{J}\left[ \psi_{2}'^{*}\psi_{1}' + \psi_{1}'^{*}\psi_{2}'   \right] + \frac{5i\sqrt{\lambda}}{2J}\left[ \psi_{1}^{*}\psi_{2}'\psi_{2}\psi_{1}'^{*}\psi_{1}\psi_{2}^{*} \right. \notag \\
&\left. + \psi_{2}'^{*}\psi_{1}\psi_{2}\psi_{1}^{*}\psi_{1}'\psi_{2}^{*}
\right] - \frac{4i\lambda g_{2}}{J^{2}}\left[ \psi_{2}'^{*}\psi_{2}\psi_{1}'^{*}\psi_{1} + \psi_{1}'^{*}\psi_{1}'\psi_{2}^{*}\psi_{2}' - \psi_{2}'^{*}\psi_{2}'\psi_{1}^{*}\psi_{1} -\psi_{1}'^{*}\psi_{1}\psi_{2}'^{*}\psi_{2}' \right] \notag \\
&+ \frac{16i (\lambda)^{3/2}(g_{2})^{2}}{J^{3}} 
\left[ \psi_{1}'^{*}\psi_{2}\psi_{2}'^{*}\psi_{1}^{*}\psi_{2}'\psi_{1}' - \psi_{1}'^{*}\psi_{2}^{*}\psi_{1}\psi_{2}'\psi_{2}'^{*}\psi_{1}' \right] + \frac{9i \lambda}{4J^{2}}\left[ \psi_{2}'\psi_{1}^{*}\psi_{1}'\psi_{2}^{*}\psi_{1}'^{*}\psi_{1} \right. \notag \\
&\left. +\psi_{2}'\psi_{1}^{*}\psi_{1}'\psi_{2}^{*}\psi_{2}'^{*}\psi_{2}
+ \psi_{2}^{*}\psi_{2}'\psi_{2}\psi_{1}'^{*}\psi_{1}\psi_{2}'^{*} + \psi_{1}^{*}\psi_{1}'\psi_{2}\psi_{1}'^{*}\psi_{1}\psi_{2}'^{*} \right]. \label{identity_4}
\end{align}
The constraint \eqref{zero_anomaly} corresponding to the zero anomalous term can be proved by first writing the left hand side as:
\begin{equation}
	\phi^{11} + \phi^{22} = (l_{3}^{2} + l_{4}^{2})C_{1} + (2il_{3}l_{4})C_{2}, \label{identity_5}
\end{equation}
where
\begin{align}
	C_{1} &= \partial_{1}(\alpha_{0}\psi_{2}^{*})(2\alpha_{0}\gamma_{\tau}\psi_{2}) + (2\alpha_{0}\gamma_{\tau}\psi_{2}^{*})\partial_{1}(\alpha_{0}\psi_{2}) + \partial_{0}(\alpha_{0}\psi_{2})(2\alpha_{0}\gamma_{\sigma}\psi_{2}^{*}) +  (2\alpha_{0}\gamma_{\sigma}\psi_{2})\partial_{0}(\alpha_{0}\psi_{2}^{*}) \notag \\
	&+ \partial_{1}(\alpha_{0}\psi_{1})(2\alpha_{0}\gamma_{\tau}\psi_{1}^{*}) + (2\alpha_{0}\gamma_{\tau}\psi_{1})\partial_{1}(\alpha_{0}\psi_{1}^{*}) + 
	\partial_{0}(\alpha_{0}\psi_{1}^{*})(2\alpha_{0}\gamma_{\sigma}\psi_{1}) + (2\alpha_{0}\gamma_{\sigma}\psi_{1}^{*})\partial_{0}(\alpha_{0}\psi_{1}) \notag \\
	&+ \partial_{0}(\alpha_{0}\psi_{2})\partial_{1}(\alpha_{0}\psi_{2}^{*}) + \partial_{0}(\alpha_{0}\psi_{2}^{*})\partial_{1}(\alpha_{0}\psi_{2}) +\partial_{0}(\alpha_{0}\psi_{1})\partial_{1}(\alpha_{0}\psi_{1}^{*}) + \partial_{0}(\alpha_{0}\psi_{1}^{*})\partial_{1}(\alpha_{0}\psi_{1}), \label{identity_6}
\end{align}
and
\begin{align}
	C_{2} &= (2\alpha_{0}\gamma_{\tau}\psi_{2})\partial_{1}(\alpha_{0}\psi_{1}^{*}) + (2\alpha_{0}\gamma_{\tau}\psi_{2}^{*})\partial_{1}(\alpha_{0}\psi_{1}) + \partial_{0}(\alpha_{0}\psi_{2})(2\alpha_{0}\gamma_{\sigma}\psi_{1}^{*}) +  \partial_{0}(\alpha_{0}\psi_{2}^{*}) (2\alpha_{0}\gamma_{\sigma}\psi_{1}) \notag \\
	&+ (2\alpha_{0}\gamma_{\tau}\psi_{1})\partial_{1}(\alpha_{0}\psi_{2}^{*}) + (2\alpha_{0}\gamma_{\tau}\psi_{1}^{*})\partial_{1}(\alpha_{0}\psi_{2}) + 
	\partial_{0}(\alpha_{0}\psi_{1})(2\alpha_{0}\gamma_{\sigma}\psi_{2}^{*}) + \partial_{0}(\alpha_{0}\psi_{1}^{*}) (2\alpha_{0}\gamma_{\sigma}\psi_{2}) \notag \\
	&+ \partial_{1}(\alpha_{0}\psi_{1}^{*})\partial_{0}(\alpha_{0}\psi_{2}) +  \partial_{0}(\alpha_{0}\psi_{2}^{*})\partial_{1}(\alpha_{0}\psi_{1}) +\partial_{0}(\alpha_{0}\psi_{1})\partial_{1}(\alpha_{0}\psi_{2}^{*}) + \partial_{1}(\alpha_{0}\psi_{2})\partial_{0}(\alpha_{0}\psi_{1}^{*}). \label{identity_7}
\end{align}
Then, one simply shows that $C_{1} = 0$, and $C_{2}=0$.

\section{Matrix elements of $\Omega_{ij}$}\label{app_omega_elements}
In this appendix we give the explicit formulas for the matrix elements of $\Omega(x)$. For completeness, we give the explicit dependence of these functions on the coupling constants $g_{2}$ and $g_{3}$ (see footnote \ref{fn:dbsec}).
\begin{align}
	\Omega_{11} &= \frac{i \sqrt{\lambda} \; g_2}{2 J^2} \left( \chi_3 \chi_4' - \chi_4 \chi_3' \right) + \frac{i \sqrt{\lambda} \;g_3}{4 J^3} \chi_2\chi_3\chi_4\chi_4', \displaybreak[3] \\
	\Omega_{12} &= \frac{i \sqrt{\lambda} \;g_3}{8J^3} \left( \chi_2 \chi_3 \chi_4 \chi_3' - \chi_1 \chi_3 \chi_4 \chi_4' \right), \displaybreak[3] \\
	\Omega_{13} &= -i + \frac{i \sqrt{\lambda} \;g_2}{2 J^2} \left( \chi_2 \chi_3' - \chi_4 \chi_1' \right) + \frac{i \sqrt{\lambda} \; g_3}{8 J^3} \left( \chi_2 \chi_3 \chi_4 \chi_2' + \chi_1 \chi_2 \chi_4 \chi_4' \right), \displaybreak[3] \\
	\Omega_{14} &= \frac{i \sqrt{\lambda} \; g_2}{2 J^2} \left( \chi_2 \chi_4' + \chi_3 \chi_1' \right) \nonumber \\
			      &+ \frac{i \sqrt{\lambda} \;g_3}{8J^3} \left( 2 \chi_2 \chi_3 \chi_4 \chi_1' - 2 \chi_1 \chi_2 \chi_3 \chi_4' - \chi_1 \chi_3 \chi_4 \chi_2' + \chi_1 \chi_2 \chi_4 \chi_3' \right), \displaybreak[3]\\
	\Omega_{22} &= \frac{i \sqrt{\lambda}\; g_2}{2 J^2} \left( \chi_3 \chi_4' - \chi_4 \chi_3' \right) - \frac{i \sqrt{\lambda} \;g_3}{4J^3} \chi_1 \chi_3 \chi_4 \chi_3', \displaybreak[3] \\
	\Omega_{23} &= \frac{i \sqrt{\lambda} \; g_2}{2 J^2} \left( -\chi_1 \chi_3' - \chi_4 \chi_2' \right) \nonumber \\
			      &+ \frac{i \sqrt{\lambda} \; g_3}{8 J^3} \left( -2 \chi_1 \chi_3 \chi_4 \chi_2' +2 \chi_1 \chi_2 \chi_4 \chi_3' + \chi_2 \chi_3 \chi_4 \chi_1' - \chi_1 \chi_2 \chi_3 \chi_4' \right), \displaybreak[3] \\
	\Omega_{24} &= -i + \frac{i \sqrt{\lambda}\; g_2}{2 J^2} \left( -\chi_1 \chi_4' + \chi_3 \chi_2' \right) + \frac{i \sqrt{\lambda} \; g_3}{8J^3} \left( -\chi_1 \chi_3 \chi_4 \chi_1' - \chi_1 \chi_2 \chi_3 \chi_3' \right), \displaybreak[3] \\
	\Omega_{33} &= -\frac{i \sqrt{\lambda}\;g_2}{2 J^2} \left( \chi_1 \chi_2' - \chi_2 \chi_1' \right) + \frac{i \sqrt{\lambda}}{4 J^3} \chi_1 \chi_2 \chi_4 \chi_2', \displaybreak[3]\\
	\Omega_{34} &= \frac{i \sqrt{\lambda}\; g_3}{8J^3} \left( \chi_1 \chi_2 \chi_4 \chi_1' - \chi_1 \chi_2 \chi_3 \chi_2' \right),\displaybreak[3] \\
	\Omega_{44} &= \frac{i \sqrt{\lambda} \; g_2}{2 J^2} \left(\chi_2 \chi_1' - \chi_1 \chi_2' \right) - \frac{i \sqrt{\lambda} \; g_3}{4 J^3} \chi_1 \chi_2 \chi_3 \chi_1'.
\end{align}
We note that $\Omega(x)$ is a symmetric matrix, 
\begin{equation}
	\Omega_{ij} = \Omega_{ji}.
\end{equation}
Moreover, its elements satisfy the following properties under complex conjugation:
\begin{align}
	\Omega_{11}^* = - \Omega_{33}  , \quad \Omega_{12}^* = - \Omega_{34} , \quad \Omega_{13}^* = - \Omega_{13} , \quad \Omega_{14}^* = - \Omega_{23}  , \quad \Omega_{22}^* = - \Omega_{44} , \quad \Omega_{22}^* = - \Omega_{44}.
\end{align}

\section{The Dirac structure}\label{app_poisson_structure}
In this appendix we present the Dirac structure of the $AAF$ model. For completeness, here we also give the explicit dependence on the coupling constants $g_{2}$ and $g_{3}$ (see footnote \ref{fn:dbsec}).

\begin{align}
(\Omega^{\scriptscriptstyle{-1}})_{11} &=  \frac{i \sqrt{\lambda}}{2J^2} g_2 \left( -\chi_1 \chi_2' + \chi_2 \chi_1'\right) + \frac{i \sqrt{\lambda}}{4J^3}g_3 \chi_1 \chi_2\chi_4 \chi_2'  + \\
	&+ \frac{i\lambda}{2J^4}g_2^2 \left( \chi_1\chi_2\chi_2'\chi_3' + \chi_1 \chi_4\chi_1' \chi_2' \right) + \nonumber \\
	&+ \frac{i \lambda^{\frac{3}{2}}}{4J^6}g_2^3 \left( -\chi_1\chi_2 \chi_3 \chi_1' \chi_2' \chi_4' - 3 \chi_1\chi_2 \chi_4 \chi_1' \chi_2' \chi_3' \right) , \nonumber \displaybreak[3] \\
(\Omega^{\scriptscriptstyle{-1}})_{22} &= \frac{i \sqrt{\lambda}}{2J^2} g_2 \left( -\chi_1 \chi_2' + \chi_2 \chi_1'\right) - \frac{i \sqrt{\lambda}}{4J^3}g_3 \chi_1 \chi_2\chi_3 \chi_1' + \\
	&+ \frac{i\lambda}{2J^4}g_2^2 \left(- \chi_1\chi_2\chi_1'\chi_4' - \chi_2 \chi_3\chi_1' \chi_2' \right) + \nonumber \\
	&+ \frac{i \lambda^{\frac{3}{2}}}{4J^6}g_2^3 \left( 3 \chi_1\chi_2 \chi_3 \chi_1' \chi_2' \chi_4' + \chi_1\chi_2 \chi_4 \chi_1' \chi_2' \chi_3' \right) , \nonumber \displaybreak[3] \\
(\Omega^{\scriptscriptstyle{-1}})_{12} &=  \frac{i \sqrt{\lambda}}{8 J^3} g_3 \left( -\chi_1 \chi_2 \chi_3 \chi_2' + \chi_1\chi_2 \chi_4 \chi_1' \right) + \\
	&+ \frac{i \lambda}{4J^4}g_2^2 \left( -\chi_1 \chi_3 \chi_1' \chi_2' + \chi_1\chi_2 \chi_2' \chi_4' - \chi_1\chi_2\chi_1' \chi_3' + \chi_2\chi_4\chi_1'\chi_2' \right)- \nonumber\\
	&- \frac{i\lambda}{2J^5}g_2g_3 \chi_1\chi_2\chi_3\chi_4\chi_1'\chi_2' + \frac{i \lambda^{\frac{3}{2}}}{2 J^6}g_2^3 \left( \chi_1\chi_2\chi_3\chi_1'\chi_2'\chi_3' -  \chi_1\chi_2\chi_4\chi_1'\chi_2'\chi_4' \right), \nonumber
	\displaybreak[3] \\
(\Omega^{\scriptscriptstyle{-1}})_{14} &= - \frac{i \sqrt{\lambda}}{2J^2} g_2\left( \chi_1 \chi_3' + \chi_4 \chi_2'\right) + \\
	&+ \frac{i \sqrt{\lambda}}{8J^3} g_3 \left( - \chi_1\chi_2\chi_3\chi_4' + \chi_2 \chi_3 \chi_4 \chi_1' + 2 \chi_1\chi_2 \chi_4 \chi_3' - 2\chi_1\chi_3\chi_4\chi_2' \right) \nonumber + \\
	&+ \frac{i {\lambda}}{4 J^4}g_2^2 \left( -\chi_1 \chi_3\chi_2' \chi_3' + \chi_1\chi_4\chi_1' \chi_3' + \chi_1\chi_4\chi_2' \chi_4' - \chi_2 \chi_4\chi_2'\chi_3' \right) - \nonumber \\
	&-\frac{i {\lambda}^{\frac{3}{2}}}{4J^6}g_2^3 \left( \chi_1\chi_2\chi_3 \chi_1' \chi_3' \chi_4' + \chi_2\chi_3 \chi_4 \chi_1' \chi_2' \chi_4' \right), \nonumber \displaybreak[3] \\
(\Omega^{\scriptscriptstyle{-1}})_{23} &=  \frac{i \sqrt{\lambda}}{2J^2} g_2\left( \chi_3 \chi_1' + \chi_2 \chi_4'\right) + \\
	&+ \frac{i \sqrt{\lambda}}{8J^3} g_3 \left(  \chi_1\chi_2\chi_4 \chi_3' - \chi_1 \chi_3 \chi_4 \chi_2' - 2 \chi_1\chi_2 \chi_3 \chi_4' + 2\chi_2\chi_3\chi_4\chi_1' \right) \nonumber + \\
	&+ \frac{i {\lambda}}{4 J^4}g_2^2 \left( -\chi_1 \chi_3\chi_1'\chi_4' + \chi_2\chi_3\chi_1'\chi_3' + \chi_2\chi_3\chi_2'\chi_4' - \chi_2 \chi_4\chi_1'\chi_4' \right) + \nonumber \\
	&+\frac{i {\lambda}^{\frac{3}{2}}}{4J^6}g_2^3 \left( \chi_1\chi_2\chi_4 \chi_2' \chi_3' \chi_4' + \chi_1\chi_3 \chi_4 \chi_1' \chi_2' \chi_3' \right), \nonumber 
	\displaybreak[3] \\
(\Omega^{\scriptscriptstyle{-1}})_{13} &= i +\frac{i\sqrt{\lambda}}{2J^2}g_2 \left( \chi_2 \chi_3' - \chi_4\chi_1' \right) + \\
	&+ \frac{i \sqrt{\lambda}}{8J^3}g_3 \chi_2\chi_4 \left(\chi_1\chi_4' - \chi_3\chi_2'\right) + \nonumber \\
	&+ \frac{i\lambda}{4J^4}g_2^2 \left( \chi_1\chi_2\chi_3'\chi_4' - \chi_1\chi_3 \chi_1' \chi_3' + \chi_1\chi_3\chi_2'\chi_4' - \chi_1\chi_4\chi_2'\chi_3' \right. - \nonumber\\
	&- \left. \chi_2 \chi_3\chi_1' \chi_4' - \chi_2\chi_4\chi_1'\chi_3' - \chi_2\chi_4 \chi_2'\chi_4' + \chi_3\chi_4\chi_1'\chi_2'\right) + \nonumber \\
	&+\frac{i \lambda}{4 J^5}g_2g_3 \chi_1\chi_2\chi_3\chi_4 \left( - \chi_1'\chi_3' +2\chi_2'\chi_4'  \right) + \nonumber \\
	&+\frac{i{\lambda}^{\frac{3}{2}}}{4J^6}g_2^3\left( - \chi_1\chi_2\chi_4\chi_1'\chi_3'\chi_4' + \chi_2\chi_3\chi_4\chi_1'\chi_2'\chi_3' + 2 \chi_1\chi_2\chi_3\chi_2'\chi_3'\chi_4' - 2 \chi_1 \chi_3\chi_4\chi_1'\chi_2'\chi_4' \right) - \nonumber \\
	&- \frac{3i{\lambda}^2 g_{2}^4}{4 J^8} \chi_1\chi_2\chi_3\chi_4\chi_1'\chi_2'\chi_3'\chi_4', \nonumber \displaybreak[3] \\
(\Omega^{\scriptscriptstyle{-1}})_{24} &= i -\frac{i\sqrt{\lambda}}{2J^2}g_2 \left( \chi_1 \chi_4' - \chi_3\chi_2' \right) + \\
	&+ \frac{i \sqrt{\lambda}}{8J^3}g_3 \chi_1\chi_3 \left(\chi_2\chi_3' - \chi_4\chi_1'\right) + \nonumber \\
	&+ \frac{i\lambda}{4J^4}g_2^2 \left( \chi_1\chi_2\chi_3'\chi_4' - \chi_1\chi_3 \chi_1' \chi_3' - \chi_1\chi_3\chi_2'\chi_4' - \chi_1\chi_4\chi_2'\chi_3' \right. - \nonumber\\
	&- \left. \chi_2 \chi_3\chi_1' \chi_4' + \chi_2\chi_4\chi_1'\chi_3' - \chi_2\chi_4 \chi_2'\chi_4' + \chi_3\chi_4\chi_1'\chi_2'\right) + \nonumber \\
	&+\frac{i \lambda}{4 J^5}g_2g_3 \chi_1\chi_2\chi_3\chi_4 \left( - 2 \chi_1'\chi_3' +\chi_2'\chi_4'  \right) + \nonumber \\
	&+\frac{i{\lambda}^{\frac{3}{2}}}{4J^6}g_2^3\left( \chi_1\chi_2\chi_3\chi_2'\chi_3'\chi_4' - \chi_1\chi_3\chi_4\chi_1'\chi_2'\chi_4' - 2 \chi_1\chi_2\chi_4\chi_1'\chi_3'\chi_4' + 2 \chi_2 \chi_3\chi_4\chi_1'\chi_2'\chi_3' \right) - \nonumber \\
	&- \frac{3i{\lambda}^2 g_{2}^{4}}{4 J^8} \chi_1\chi_2\chi_3\chi_4\chi_1'\chi_2'\chi_3'\chi_4', \nonumber \displaybreak[3] \\
	(\Omega^{\scriptscriptstyle{-1}})_{33} &=  \frac{i \sqrt{\lambda}}{2J^2} g_2 \left( -\chi_4'\chi_3  + \chi_3'\chi_4 \right) + \frac{i \sqrt{\lambda}}{4J^3}g_3 \chi_4'\chi_2\chi_4\chi_3   + \\
	&+ \frac{i\lambda}{2J^4}g_2^2 \left( \chi_1' \chi_4' \chi_4 \chi_3 + \chi_4' \chi_3' \chi_2\chi_3  \right) + \nonumber \\
	&+ \frac{i \lambda^{\frac{3}{2}}}{4J^6}g_2^3 \left( - \chi_2'\chi_4'\chi_3'\chi_1\chi_4\chi_3     - 3 \chi_1'\chi_4'\chi_3'\chi_2\chi_4\chi_3     \right) , \nonumber \displaybreak[3] \\
(\Omega^{\scriptscriptstyle{-1}})_{44} &= \frac{i \sqrt{\lambda}}{2J^2} g_2 \left( -\chi_4'\chi_3  + \chi_3'\chi_4 \right) - \frac{i \sqrt{\lambda}}{4J^3}g_3  \chi_3'\chi_1 \chi_4\chi_3 + \\
	&+ \frac{i\lambda}{2J^4}g_2^2 \left(- \chi_2'\chi_3'\chi_4\chi_3 - \chi_4' \chi_3'\chi_1 \chi_4 \right) + \nonumber \\
	&+ \frac{i \lambda^{\frac{3}{2}}}{4J^6}g_2^3 \left( 3 \chi_2'\chi_4'\chi_3' \chi_1 \chi_4 \chi_3 +  \chi_1'\chi_4' \chi_3' \chi_2 \chi_4 \chi_3 \right) , \nonumber \displaybreak[3] \\
(\Omega^{\scriptscriptstyle{-1}})_{34} &=  \frac{i \sqrt{\lambda}}{8 J^3} g_3 \left( -\chi_4' \chi_1 \chi_4 \chi_3 + \chi_3'\chi_2 \chi_4 \chi_3 \right) + \\
	&+ \frac{i \lambda}{4J^4}g_2^2 \left( -\chi_4' \chi_3' \chi_1 \chi_3 + \chi_2'\chi_4' \chi_4 \chi_3 - \chi_1'\chi_3'\chi_4 \chi_3 + \chi_4'\chi_3'\chi_2\chi_4 \right)- \nonumber\\
	&- \frac{i\lambda}{2J^5}g_2g_3 \chi_4'\chi_3'\chi_2\chi_1\chi_4\chi_3 + \frac{i \lambda^{\frac{3}{2}}}{2 J^6}g_2^3 \left( \chi_1'\chi_4'\chi_3'\chi_1\chi_4\chi_3 -  \chi_2'\chi_4'\chi_3'\chi_2\chi_4\chi_3 \right), \nonumber
\end{align}
We note that the matrix $\Omega^{\scriptscriptstyle{-1}}(x)$ is also symmetric:
\begin{equation}
(\Omega^{\scriptscriptstyle{-1}})_{ij}=(\Omega^{\scriptscriptstyle{-1}})_{ji}.
\end{equation}
Furthermore, its elements satisfy the following properties under involution:
\begin{align}
	{(\Omega^{\scriptscriptstyle{-1}})^*_{33}} &=-(\Omega^{\scriptscriptstyle{-1}})_{11}, \quad {(\Omega^{\scriptscriptstyle{-1}})^*_{44}} =-(\Omega^{\scriptscriptstyle{-1}})_{22}, \quad{(\Omega^{\scriptscriptstyle{-1}})^*_{34}} =-(\Omega^{\scriptscriptstyle{-1}})_{12},\notag \\
{(\Omega^{\scriptscriptstyle{-1}})^*_{14}} &=-(\Omega^{\scriptscriptstyle{-1}})_{32}, \quad {(\Omega^{\scriptscriptstyle{-1}})^*_{13}} =-(\Omega^{\scriptscriptstyle{-1}})_{13}, \quad {(\Omega^{\scriptscriptstyle{-1}})^*_{24}} =-(\Omega^{\scriptscriptstyle{-1}})_{24}. \label{omega_relations}
\end{align}

Finally, we address a subtlety related to the validity of the Jacobi identity. Even when considering the simplest forms of the Jacobi identity, such as:
\begin{equation}\label{ji_problem}
	\sum_{\sigma \in \mathbb{P}_c} \left\{ \chi_1 \left( x_{\sigma(1)}\right), \left\{ \chi_1 \left( x_{\sigma(2)}\right), \chi_1 \left( x_{\sigma(3)}\right) \right\} \right\} = 0,
\end{equation}
where $\mathbb{P}_c$ stands for all cyclic permutations of (1,2,3), one arrives at meaningless expressions, since the left hand side fails to vanish. One has to remember, however, that the correct way to define the Poisson brackets in field theory is done by introducing functionals as in \eqref{functionals}:
\begin{equation}
	F(x) = \int d\xi \: c(x,\xi) \chi_1(\xi),
\end{equation}
where $c(x,\xi)$ is some smooth generalized function with some properties on the boundary. The corresponding Jacobi identity is then satisfied, provided some general conditions on the functionals (see \cite{Faddeev:1987ph, olver1986applications} for more details). 

\section{Computational details for the algebra of Lax operators}\label{app_lax_details}

In this appendix, we collect some useful formulae necessary for the derivation of the algebra of Lax operators in section \ref{sec_lax_algebra}. There we also use the following representation for the two-dimensional Dirac matrices:
\begin{align}
	\sigma^3 = \left( \begin{array}{cc}
		1 & 0 \\
		0 & -1 \end{array} \right) , \quad
	\sigma^+ = \left( \begin{array}{cc}
		0 & 1 \\
		0 & 0 \end{array} \right) , \quad
	\sigma^- = \left( \begin{array}{cc}
		0 & 0 \\
		1 & 0 \end{array} \right).
\end{align}

The functions $\xi^{(\sigma)}_j(x;\mu)$ and $\Lambda^{(\pm)}_{\sigma}(x;\mu)$ used in the decomposition \eqref{lax_decomposition} of the spacial component of the Lax connection are:
\begin{align}
	\xi^{(\sigma)}_0 &= \frac{1}{4J} \left[ - \chi_3\chi_1' + \chi_4\chi_2' - \chi_1\chi_3' + \chi_2\chi_4' \right]\\
	&- \frac{1}{4J^2}\left[\chi_2\chi_3\chi_4\chi_1' - \chi_1\chi_3\chi_4\chi_2' - \chi_1\chi_2\chi_4\chi_3' + \chi_1\chi_2\chi_3\chi_4' \right], \nonumber \displaybreak[3]\\
	\xi^{(\sigma)}_1 &= \frac{l_1}{8J} \left[ \chi_3\chi_1' +\chi_4\chi_2' +\chi_1\chi_3' +\chi_2\chi_4' \right] \\ 
	&+ \frac{il_2}{4 \sqrt{\lambda}} \left[ 2J +\frac{\sqrt{\lambda}}{2J} \left( \chi_4\chi_1' - \chi_3\chi_2' + \chi_1\chi_4' - \chi_2\chi_3' \right) +  \left(-\chi_1\chi_3 + \chi_2\chi_4 \right) \right], \nonumber \displaybreak[3]\\
	\Lambda^{(-)}_{\sigma} &= \bar{\Lambda}^1_{\sigma} - i \bar{\Lambda}^{2}_{\sigma} \\
	&= \frac{l_3}{\sqrt{J}} \left[ -\chi_2' +\frac{1}{2J} \chi_2\chi_3\chi_1' +\frac{1}{4J} \left( \chi_2\chi_4\chi_2' - \chi_1\chi_3\chi_2' \right) -\frac{1}{16J^2} \chi_1\chi_2\chi_3\chi_4\chi_2' \right] \nonumber \\
	&+ \frac{i l_4}{\sqrt{J}}  \left[ -\chi_1' -\frac{1}{2J} \chi_1\chi_2\chi_3' +\frac{1}{4J} \left( \chi_2\chi_4\chi_1' - \chi_1\chi_3\chi_1' \right) -\frac{1}{16J^2} \chi_1\chi_2\chi_3\chi_4\chi_1' \right], \nonumber \displaybreak[3]\\
	\Lambda^{(+)}_{\sigma} &= \bar{\Lambda}^1_{\sigma} + i \bar{\Lambda}^{2}_{\sigma} \\
	&= \frac{l_3}{\sqrt{J}} \left[ -\chi_4' +\frac{1}{2J} \chi_1\chi_4\chi_3' +\frac{1}{4J} \left( \chi_2\chi_4\chi_4' - \chi_1\chi_3\chi_4' \right) -\frac{1}{16J^2} \chi_1\chi_2\chi_3\chi_4\chi_4' \right] \nonumber \\
	&+ \frac{i l_4}{\sqrt{J}}  \left[ \chi_3' +\frac{1}{2J} \chi_2\chi_3\chi_4' -\frac{1}{4J} \left( \chi_2\chi_4\chi_3' - \chi_1\chi_3\chi_3' \right) +\frac{1}{16J^2} \chi_1\chi_2\chi_3\chi_4\chi_3' \right], \nonumber 
\end{align}
where we dropped the dependence on $x$ and on the spectral parameter $\mu$ to avoid cluttering.

The functions $\Gamma_{11}$, $\Gamma_{12}^{(1)}$ and $\Gamma_{12}^{(2)}$ appearing in the expressions \eqref{pb_a1a1}-\eqref{pb_a1a2} and \eqref{pb_a2a1} are:
\begin{align}
	\Gamma_{11} &= \chi_1 \left( \chi_3' \Omega^{\scriptscriptstyle{-1}}_{13} -  \chi_4' \Omega^{\scriptscriptstyle{-1}}_{23} + \chi_1' \Omega^{\scriptscriptstyle{-1}}_{33} - \chi_2' \Omega^{\scriptscriptstyle{-1}}_{44} \right) 
				+ \chi_2 \left( -\chi_3' \Omega^{\scriptscriptstyle{-1}}_{14} +\chi_4' \Omega^{\scriptscriptstyle{-1}}_{24} - \chi_1' \Omega^{\scriptscriptstyle{-1}}_{34} + \chi_2' \Omega^{\scriptscriptstyle{-1}}_{44} \right) \nonumber \\
				&+ \chi_3 \left( \chi_3' \Omega^{\scriptscriptstyle{-1}}_{11} - \chi_4' \Omega^{\scriptscriptstyle{-1}}_{12} + \chi_1' \Omega^{\scriptscriptstyle{-1}}_{13} - \chi_2' \Omega^{\scriptscriptstyle{-1}}_{14} \right)
				+ \chi_4 \left( -\chi_3' \Omega^{\scriptscriptstyle{-1}}_{12} + \chi_2' \Omega^{\scriptscriptstyle{-1}}_{22} - \chi_1' \Omega^{\scriptscriptstyle{-1}}_{23} + \chi_2' \Omega^{\scriptscriptstyle{-1}}_{24} \right), \displaybreak[3] \\
	\Gamma_{12}^{(1)} &= 2 \partial_x \left[ \chi_2\chi_3\chi_4\chi_3' \Omega_{11}^{\scriptscriptstyle{-1}} + \left( -\chi_1\chi_3\chi_4\chi_3' - \chi_2\chi_3\chi_4\chi_4' \right) \Omega_{12}^{\scriptscriptstyle{-1}}  \right. \nonumber \\
	&+ \left. \left( \chi_2\chi_3\chi_4\chi_1' - \chi_1\chi_3\chi_4\chi_2' -\chi_1\chi_2\chi_4\chi_3' + \chi_1\chi_2\chi_3\chi_4' \right) \Omega_{13}^{\scriptscriptstyle{-1}} + \chi_1\chi_3\chi_4\chi_4' \Omega_{22}^{\scriptscriptstyle{-1}} \right. \nonumber \\
	&+ \left. \left(-\chi_2\chi_3\chi_4\chi_1' +\chi_1\chi_3\chi_4\chi_2' +\chi_1\chi_2\chi_4\chi_3' - \chi_1\chi_2\chi_3\chi_4' \right)\Omega_{24}^{\scriptscriptstyle{-1}} - \chi_1\chi_2\chi_4\chi_1' \Omega_{33}^{\scriptscriptstyle{-1}} \right. \nonumber \\
	&+ \left. \left( \chi_1\chi_2\chi_3\chi_1' + \chi_1\chi_2\chi_4\chi_2' \right) \Omega_{34}^{\scriptscriptstyle{-1}} - \chi_1\chi_2\chi_3\chi_2' \Omega^{\scriptscriptstyle{-1}}_{44}\right] - 4\chi_2\chi_3\chi_3'\chi_4' \Omega_{11}^{\scriptscriptstyle{-1}} \nonumber \\
	&+ 4\left(\chi_1\chi_3\chi_3'\chi_4' + \chi_2\chi_4\chi_3'\chi_4' \right)\Omega_{12}^{\scriptscriptstyle{-1}} + 4\left(\chi_1\chi_4\chi_2'\chi_3' - \chi_2\chi_3\chi_1'\chi_4' \right) \Omega_{13}^{\scriptscriptstyle{-1}} \nonumber \\
	&+ 4\left(\chi_2\chi_3\chi_1'\chi_3' - \chi_1\chi_3\chi_2'\chi_3' + \chi_2\chi_3\chi_2'\chi_4' - \chi_2\chi_4\chi_2'\chi_3' \right)\Omega_{14}^{\scriptscriptstyle{-1}} -4\chi_1\chi_4\chi_3'\chi_4' \Omega_{22}^{\scriptscriptstyle{-1}} \nonumber \\
	&+ 4\left(\chi_1\chi_3\chi_1'\chi_4' - \chi_1\chi_4\chi_1'\chi_3' + \chi_2\chi_4\chi_1'\chi_4' - \chi_1\chi_4\chi_2'\chi_4' \right)\Omega_{23}^{\scriptscriptstyle{-1}} + 4\left(\chi_1\chi_4\chi_2'\chi_3' - \chi_2\chi_3\chi_1'\chi_4' \right)\Omega_{24}^{\scriptscriptstyle{-1}} \nonumber \\
	&-4 \chi_1\chi_4\chi_1'\chi_2' \Omega_{33}^{\scriptscriptstyle{-1}} +4 \left(\chi_1\chi_3\chi_1'\chi_2' + \chi_2\chi_4\chi_1'\chi_2' \right) \Omega_{34}^{\scriptscriptstyle{-1}} -4 \chi_2\chi_3\chi_1'\chi_2' \Omega_{44}^{\scriptscriptstyle{-1}}, \displaybreak[3] \\
	 \Gamma_{12}^{(2)} &= 4 \chi_2\chi_3\chi_4\chi_3' \Omega_{11}^{\scriptscriptstyle{-1}} - 4 \left( \chi_1\chi_3\chi_4\chi_3' + \chi_2\chi_3\chi_4\chi_4' \right)\Omega_{12}^{\scriptscriptstyle{-1}} \nonumber \\
	 &+2 \left( 2 \chi_2\chi_3\chi_4\chi_1' - \chi_1\chi_3\chi_4\chi_2' -2 \chi_1\chi_2\chi_4\chi_3' + \chi_1\chi_2\chi_3\chi_4' \right)\Omega_{13}^{\scriptscriptstyle{-1}} + 2\left( \chi_1\chi_2\chi_3\chi_3' - \chi_2\chi_3\chi_4\chi_2'\right) \Omega_{14}^{\scriptscriptstyle{-1}} \nonumber \\
	 &+4 \chi_1\chi_3\chi_4\chi_4' \Omega_{22}^{\scriptscriptstyle{-1}} + 2 \left(-\chi_1\chi_3\chi_4\chi_1' + \chi_1\chi_2\chi_4\chi_4' \right) \Omega_{23}^{\scriptscriptstyle{-1}} \nonumber \\ 
	 &+2 \left( -\chi_2\chi_3\chi_4\chi_1' + 2 \chi_1\chi_3\chi_4\chi_2' + \chi_1\chi_2\chi_4\chi_3' -2 \chi_1\chi_2\chi_3\chi_4' \right) \Omega_{24}^{\scriptscriptstyle{-1}} -4 \chi_1\chi_2\chi_4\chi_1' \Omega_{33}^{\scriptscriptstyle{-1}}\nonumber \\
	 &+4 \left(\chi_1\chi_2\chi_3\chi_1' + \chi_1\chi_2\chi_4\chi_2' \right)\Omega_{34}^{\scriptscriptstyle{-1}} -4 \chi_1\chi_2\chi_3\chi_2' \Omega_{44}^{\scriptscriptstyle{-1}}.
\end{align}
The expression that appears in \eqref{pb_A_final}, after taking into account the explicit expressions for the Dirac structure, has the form:
\begin{align}
	2 \Gamma_{11} + \frac{\Gamma_{12}^{(2)}}{J} &= 2i \left( \chi_3\chi_1' +\chi_4\chi_2' +\chi_1\chi_3' +\chi_2\chi_4'\right) \nonumber\\ 
	&+ \frac{2i}{J}\left(\chi_2\chi_3\chi_4\chi_1' + \chi_1\chi_3\chi_4\chi_2' - \chi_1\chi_2\chi_4\chi_3' - \chi_1\chi_2\chi_3\chi_4' \right) \nonumber \\
	&+ \frac{2i \sqrt{\lambda}}{J^2} \left( \chi_1\chi_3\chi_2'\chi_3' -\chi_1\chi_3\chi_1'\chi_4' + \chi_2\chi_4\chi_2'\chi_3' - \chi_2\chi_4\chi_1'\chi_4' \right) \nonumber\\
	&+ \frac{i\lambda}{J^4} \left( \chi_1\chi_2\chi_4\chi_2'\chi_3'\chi_4' + \chi_1\chi_2\chi_3 \chi_1'\chi_3'\chi_4' + \chi_2\chi_3\chi_4 \chi_1'\chi_2'\chi_4' + \chi_1\chi_3\chi_4\chi_1'\chi_2'\chi_3'\right).
\end{align}

\section{The List of $N_i^{(j)}(\mu_1,\mu_2)$ functions}\label{app_N_functions}
In this appendix, we list the functions $N_i^{(j)}(\mu_1,\mu_2)$ appearing in the matrices $N_i(x,y;\mu_1,\mu_2)$, $i=0,1,2$, which provide the non-ultralocal decomposition of the Lax algebra for the fermionic Wadati model \eqref{wadati_LL_algebra}.
\begin{align}
	N_0^{(1)}(\mu_1,\mu_2) &= \frac{2}{\sqrt{\lambda J}} l_2(\mu_2) \left[ i l_4(\mu_1) \chi_3' + l_3(\mu_1) \chi_4' \right] + \frac{2 \alpha_1(\mu_2,\mu_1)}{J^{\frac{3}{2}}}\chi_3'' - \frac{2 i \beta_1(\mu_2,\mu_1)}{J^{\frac{3}{2}}} \chi_4'', \\
	N_0^{(2)}(\mu_1,\mu_2) &= \frac{2}{\sqrt{\lambda J}} l_2(\mu_2) \left[ - i l_4(\mu_1) \chi_3' - l_3(\mu_1) \chi_4' \right] + \frac{2 \alpha_2(\mu_2,\mu_1)}{J^{\frac{3}{2}}}\chi_3'' - \frac{2 i \beta_2(\mu_2,\mu_1)}{J^{\frac{3}{2}}} \chi_4'', \\
	N_1^{(1)}(\mu_1,\mu_2) &= \frac{2}{\sqrt{\lambda J}} l_2(\mu_1) \left[ i l_4(\mu_2) \chi_3 + l_3(\mu_2) \chi_4 \right] +  \frac{\alpha_1(\mu_1,\mu_2)}{J^{\frac{3}{2}}}\chi_3' -  \frac{i \beta_1(\mu_1,\mu_2)}{J^{\frac{3}{2}}} \chi_4', \\
	N_1^{(2)}(\mu_1,\mu_2) &= \frac{2}{\sqrt{\lambda J}} l_2(\mu_2) \left[ i l_4(\mu_1) \chi_3 + l_3(\mu_1) \chi_4 \right] +  \frac{3\alpha_1(\mu_2,\mu_1)}{J^{\frac{3}{2}}}\chi_3' -  \frac{3i \beta_1(\mu_2,\mu_1)}{J^{\frac{3}{2}}} \chi_4', \\
	N_1^{(3)}(\mu_1,\mu_2) &= \frac{2}{\sqrt{\lambda J}} l_2(\mu_2) \left[ -i l_4(\mu_1) \chi_3 - l_3(\mu_1) \chi_4 \right] +  \frac{3\alpha_2(\mu_2,\mu_1)}{J^{\frac{3}{2}}}\chi_3' - \frac{3 i \beta_2(\mu_2,\mu_1)}{J^{\frac{3}{2}}} \chi_4', \\
	N_1^{(4)}(\mu_1,\mu_2) &= \frac{2}{\sqrt{\lambda J}} l_2(\mu_1) \left[ -i l_4(\mu_2) \chi_3 - l_3(\mu_2) \chi_4 \right] +  \frac{\alpha_2(\mu_1,\mu_2)}{J^{\frac{3}{2}}}\chi_3' -  \frac{i \beta_2(\mu_1,\mu_2)}{J^{\frac{3}{2}}} \chi_4', \\
	N_2^{(1)}(\mu_1,\mu_2) &= \frac{1}{J^{\frac{3}{2}}} \left[ -\alpha_1(\mu_1,\mu_2) \chi_3 + i \beta_1(\mu_1,\mu_2) \chi_4 \right], \\
	N_2^{(2)}(\mu_1,\mu_2) &=  \frac{1}{J^{\frac{3}{2}}} \left[ \alpha_1(\mu_2,\mu_1) \chi_3 - i \beta_1(\mu_2,\mu_1) \chi_4 \right], \\
	N_2^{(3)}(\mu_1,\mu_2) &=  \frac{1}{J^{\frac{3}{2}}} \left[ \alpha_2(\mu_2,\mu_1) \chi_3 - i \beta_2(\mu_2,\mu_1) \chi_4 \right], \\
	N_2^{(4)}(\mu_1,\mu_2) &=  \frac{1}{J^{\frac{3}{2}}} \left[ -\alpha_2(\mu_1,\mu_2) \chi_3 + i \beta_2(\mu_1,\mu_2) \chi_4 \right], \\ 
	N_2^{(5)}(\mu_1,\mu_2) &= \frac{8}{J} \left[- i l_3(\mu_1)l_3(\mu_2) -i l_4(\mu_1) l_4(\mu_2) \right].
\end{align}
Here $\alpha_i (\mu_1,\mu_2)$ and $\beta_i(\mu_1,\mu_2)$, $i=1,2$ are the following functions of the spectral parameters:
\begin{align}
	\alpha_1 (\mu_1,\mu_2) &= -l_2(\mu_1) l_3(\mu_2) + \left[ 2 - l_1(\mu_1) \right]l_4(\mu_2),\\
	\alpha_2(\mu_1,\mu_2) &= l_2(\mu_1) l_3(\mu_2) +\left[ 2 +l_1(\mu_1) \right] l_4(\mu_2),\\
	\beta_1(\mu_1,\mu_2) &= l_2(\mu_1) l_4(\mu_2) + \left[ 2 +l_1(\mu_1) \right]l_3(\mu_2),\\
	\beta_2(\mu_1,\mu_2) &= -l_2(\mu_1) l_4(\mu_2) + \left[ 2 -l_1(\mu_1) \right]l_3(\mu_2).
\end{align}

\section*{Acknowledgment} A.M. would like to thank A. Pinzul for useful discussions. The work of A.M. was partially supported by CAPES. The work of G.W. was supported by the FAPESP grant No. 2011/20242-3.
\newpage
\bibliographystyle{utphys} 
\bibliography{r_matrix_AAF_final}

\end{document}